\documentclass[Journal,twocolumn]{IEEEtran}
\IEEEoverridecommandlockouts
\usepackage{color}
\usepackage{graphicx}
\usepackage{amsmath}
\usepackage{amssymb}
\usepackage{algorithm}
\usepackage{algorithmic}
\usepackage{amsmath}
\usepackage{multirow}
\usepackage{booktabs}
\usepackage{array}
\usepackage{amsthm}
\usepackage{stfloats}
\usepackage{caption}
\usepackage{subfigure}
\usepackage{bm}
\usepackage{booktabs}
\usepackage{setspace}
\usepackage{url}
\usepackage{arydshln}
\usepackage{flushend}

\newcommand{\be}{\begin{equation}}
\newcommand{\ee}{\end{equation}}

\newcommand{\non}{\nonumber}

\renewcommand{\thesubsubsection}{\thesubsection.\arabic{subsubsection}}

\renewcommand{\baselinestretch}{1}
\allowdisplaybreaks[4]

\title{ 
Joint Array Partitioning and Beamforming Designs in ISAC Systems: A Bayesian CRB Perspective
\thanks{Part of this paper has been presented in Asilomar Conference on Signals Systems and Computers, 2024 \cite{RLiu-Asilomar-2024}.}
\thanks{R. Liu and A. L. Swindlehurst are with the Center for Pervasive Communications and Computing, University of California, Irvine, CA 92697, USA (e-mail: rangl2@uci.edu; swindle@uci.edu).}
\thanks{M. Li is with the School of Information and Communication Engineering, Dalian University of Technology, Dalian 116024, China (e-mail: mli@dlut.edu.cn).}
}

\author{Rang Liu,~\IEEEmembership{Member,~IEEE,}
        Ming Li,~\IEEEmembership{Senior Member,~IEEE,}
        and A. Lee Swindlehurst,~\IEEEmembership{Fellow,~IEEE}}

\begin{document}

\maketitle
\pagestyle{empty}
\thispagestyle{empty}

\begin{abstract}
Integrated sensing and communication (ISAC) has emerged as a promising paradigm for next-generation (6G) wireless networks, unifying radar sensing and communication on a shared hardware platform. This paper proposes a dynamic array partitioning framework for monostatic ISAC systems to fully exploit available spatial degrees of freedom (DoFs) and reconfigurable antenna topologies, enhancing sensing performance in complex scenarios. We first establish a theoretical foundation for our work by deriving Bayesian Cram\'{e}r-Rao bounds (BCRBs) under prior distribution constraints for heterogeneous target models, encompassing both point-like and extended targets. Building on this, we formulate a joint optimization framework for transmit beamforming and dynamic array partitioning to minimize the derived BCRBs for direction-of-arrival (DOA) estimation. The optimization problem incorporates practical constraints, including multi-user communication signal-to-interference-plus-noise ratio (SINR) requirements, transmit power budgets, and array partitioning feasibility conditions. To address the non-convexity of the problem, we develop an efficient alternating optimization algorithm combining the alternating direction method of multipliers (ADMM) with semi-definite relaxation (SDR). We also design novel maximum a posteriori (MAP) DOA estimation algorithms specifically adapted to the statistical characteristics of each target model. Extensive simulations illustrate the superiority of the proposed dynamic partitioning strategy over conventional fixed-array architectures across diverse system configurations. 
\end{abstract}

\section{Introduction}

Integrated sensing and communication (ISAC) has emerged as a foundational technology for sixth-generation (6G) wireless systems, unifying radar sensing and wireless communications within a shared framework to optimize spectral and hardware resource utilization. By harmonizing these dual functionalities, ISAC enhances spectral efficiency, reduces infrastructure costs, and supports emerging applications requiring simultaneous data transmission and environmental perception \cite{FLiu-TCOM-2020}-\cite{FLiu-SPM-2023}.
Multiple-input multiple-output (MIMO) technology plays a pivotal role in ISAC systems, leveraging spatial degrees of freedom (DoFs) to concurrently enhance communication and sensing performance. For communications, MIMO enables spatial multiplexing for high-throughput transmissions, spatial diversity for robust connectivity in multipath environments, and beamforming gains for improved spectral efficiency. For sensing, MIMO facilitates superior target detection and estimation through beamforming, space-time adaptive processing, and virtual aperture expansion, etc. However, the shared use of antenna resources in MIMO-ISAC systems introduces an inherent trade-off between communication and sensing performance, necessitating advanced beamforming techniques to balance the conflicting requirements \cite{JZhang-JSTSP-2021,ALiu-CST-2022}.

Recent developments in MIMO-ISAC beamforming designs 
have demonstrated favorable trade-offs for various performance metrics including signal-to-interference-plus-noise ratio (SINR) \cite{XLiu-TSP-2020}-\cite{ZXiao-Tcom-2025} or achievable rate \cite{ZRen-TWC-2024} for communication, and radar SINR \cite{RLiu-TWC-2024, ZXiao-Tcom-2025, RLiu-JSAC-2022, RLiu-JSTSP-2022}, beampattern similarity \cite{XLiu-TSP-2020, ZRen-Tcom-2023}, or the Cram\'{e}r-Rao bound (CRB) on estimation error \cite{FLiu-TSP-2021, RLiu-TWC-2024, XSong-TWC-2024, HHua-TWC-2024} for sensing.  Despite these advances, a fundamental limitation persists: existing designs predominantly employ fixed transmit-receive array configurations. While such architectures offer simplicity and ease of implementation, they inherently restrict the system’s ability to fully exploit available spatial DoFs. This rigidity not only limits adaptability in dynamic environments, but also degrades the achievable performance-complexity trade-off. To overcome these limitations, antenna selection has emerged as a promising solution for adaptive spatial resource allocation.

Antenna selection is a cost-effective and low-complexity strategy that retains many of the advantages of MIMO systems while reducing hardware requirements \cite{SSanayei-CM-2004}, making it attractive for both standalone MIMO communication and radar systems. For MIMO communications, the intelligent selection of a subset of antennas for transmission and reception enhances energy efficiency \cite{HLi-Tcom-2014} and achieves near-optimal capacity \cite{YGao-TSP-2018} without the need for fully populated radio frequency (RF) chains. Similarly, for MIMO radar, an optimally chosen set of transmit antennas improves waveform design, shaping the transmit beampattern \cite{WFan-TSP-2024}, minimizing the CRB for parameter estimation \cite{MArash-TWC-2023}, and maximizing radar SINR for target detection \cite{HHuang-TSP-2023}. In addition to transmit-side selection, joint selection of both transmit and receive antennas can further enhance system efficiency \cite{IVan-EUSIPCO-2024}.

Building on the proven efficacy of antenna selection in standalone communication and radar systems, recent efforts have extended this paradigm to ISAC architectures. In particular, \cite{IValiulahi-WCL-2022}-\cite{MPalaiologos-ICC-2024} focused on the transmit antenna selection while the receive array remains fixed. Similarly, receive antenna selection has been investigated in \cite{IVan-SAM-2024}, wherein a predefined subset of receive antennas is allocated for either sensing or information decoding.  Additionally, joint transmit-receive antenna selection has been explored in \cite{RSankar-SAM-2024}, where a fixed number of antennas are selected from both arrays to enhance overall system performance. While these approaches highlight the advantages of dynamic antenna selection, they share a fundamental limitation: Only a small and predetermined number of antennas are activated for transmission or reception, leaving a significant portion of the array underutilized. This restriction can result in performance degradation, particularly in scenarios with stringent sensing and communication requirements.

To overcome the inherent limitations of antenna selection-based methods, antenna array partitioning has emerged as a more flexible and superior alternative, enabling dynamic reconfiguration of the entire antenna array into transmit and receive subarrays. Unlike traditional approaches that activate only a small and fixed subset of antennas, array partitioning leverages all available elements in a reconfigurable array topology to ensure a more adaptive and resource-efficient allocation for both sensing and communication. In our previous work \cite{RLiu-TWC-2025}, we investigated joint array partitioning and beamforming design for ISAC systems, where the direction-of-arrival (DOA) estimation accuracy is optimized under communication SINR requirements and a given power budget. Simulation results demonstrated that the proposed dynamic array partitioning scheme significantly outperforms fixed partitioning methods. However, this prior study focused on scenarios with a single point-like target, which fail to capture the multifaceted challenges of practical ISAC deployments that must account for multiple point-like targets, or more general ``extended'' targets with a spatially distributed response. Such scenarios pose additional challenges for target parameter estimation and resource management. Furthermore, most existing studies optimize the CRB under idealized assumptions of perfectly known target parameters.  
To bridge this gap, a Bayesian CRB framework, which explicitly accounts for statistical uncertainties in the target parameters, is critical for designing robust ISAC systems that align with practical operational conditions.

Motivated by these issues, in this work we investigate joint array partitioning and beamforming designs for more practical and complex ISAC scenarios, developing a robust framework that efficiently manages spatial resources while jointly optimizing sensing and communication performance. The main contributions of this paper are summarized below.
\begin{itemize}
    \item We establish an ISAC model for simultaneous multiple point-like target sensing and multi-user communication with a dynamically partitioned array architecture. Based on this model, we derive the closed-form Bayesian CRB (BCRB) for target parameter estimation, providing fundamental theoretical insights. We then formulate a joint array partitioning and beamforming optimization framework to minimize the BCRB for target DOA estimation while satisfying multiuser communication SINR requirements, the transmit power budget, and inherent constraints on the array partitioning.

    \item To solve the resulting complicated problem with fractional quadratic terms and binary integer variables, we employ dedicated transformations and classical algorithmic frameworks to decompose the problem into tractable subproblems by leveraging the Schur complement and alternative direction method of multipliers (ADMM). We iteratively solve these subproblems with the aid of semi-definite relaxation (SDR). 
    
    \item We generalize our Bayesian framework to extended target sensing, where the target is characterized by its central angle and angular spread. We formulate an optimization problem to minimize the BCRB for angle estimation under constraints similar to those for the previous problem. To accommodate the extended target model, the previous optimization framework is adapted to effectively handle the additional complexity introduced by extended target representations.
    
    \item For both sensing scenarios, we develop a joint maximum a posteriori (MAP) estimation algorithm that simultaneously estimates all target parameters by maximizing the posterior distribution. Unlike conventional methods, it leverages parameter correlations and integrates prior knowledge with real-time data, enhancing accuracy and robustness, especially in low-SNR environments. This approach mitigates error propagation and offers a more efficient and reliable solution for complex sensing tasks.
    
    \item We conduct comprehensive numerical simulations to evaluate the proposed method against fixed array partitioning schemes. The results demonstrate that our dynamic partitioning approach significantly outperforms conventional methods, achieving lower root-BCRBs and reduced root-mean-squared errors (RMSEs) for various system settings.
\end{itemize}

\emph{Notation}: Boldface lower-case and upper-case letters indicate column vectors and matrices, respectively. The symbols
$(\cdot)^*$, $(\cdot)^T$, $(\cdot)^H$, and $(\cdot)^{-1}$ denote the conjugate, transpose, transpose-conjugate, and inverse operations, respectively. The space of real and imaginary numbers is respectively represented by $\mathbb{C}$ and $\mathbb{R}$. The operators
$| a |$, $\| \mathbf{a} \|$, and $\| \mathbf{A} \|_F$ represent the magnitude of a scalar $a$, the norm of a vector $\mathbf{a}$, and the Frobenius norm of a matrix $\mathbf{A}$, respectively. Statistical expectation is denoted by
$\mathbb{E}\{\cdot\}$, $\text{Tr}\{\mathbf{A}\}$ takes the trace of the matrix $\mathbf{A}$, and $\text{diag}\{\mathbf{a}\}$ indicates the diagonal matrix whose diagonal elements are taken from $\mathbf{a}$. The real part of a complex number is given by $\Re\{\cdot\}$, and  
$\mathbb{S}_N^+$ represents the set of all $N$-dimensional complex positive semidefinite matrices.
Finally, we let $\mathbf{A}(i,j)$ denote the element in the $i$-th row and the $j$-th column of matrix $\mathbf{A}$, and $\mathbf{a}(i)$ denote the $i$-th element of vector $\mathbf{a}$.

\section{System Model and Problem Formulation for Multiple Point-Like Targets}

\begin{figure}[t]	
\centering
\includegraphics[width = 0.92\linewidth]{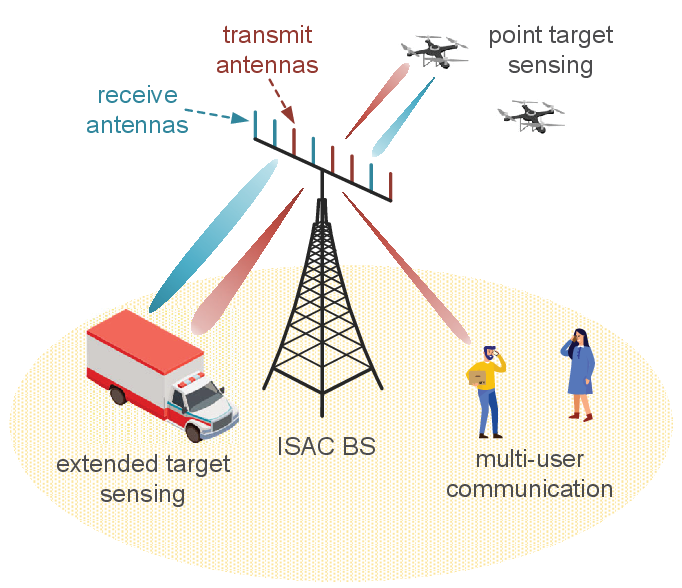}
\caption{A monostatic ISAC system with a dynamically partitioned array architecture. (red: transmit antennas and beamforming, blue: receive antennas and beamforming)}\label{fig:system}\vspace{-0.3 cm}
\end{figure}

We consider a monostatic ISAC system in which a BS is equipped with a uniform linear array (ULA) consisting of $N$ elements with half-wavelength spacing. The dual-functional BS simultaneously serves $K$ single-antenna users while performing target sensing. In this section, we address scenarios involving  multiple point-like targets, such as the drones shown in Fig.~\ref{fig:system}. Extended targets with a non-negligible angular extent, like the vehicle in Fig.~\ref{fig:system}, will be discussed in a subsequent section. These scenarios introduce increased complexity compared to conventional sensing models, necessitating sophisticated designs that dynamically and strategically optimize the allocation of available resources to achieve an optimal balance between communication and sensing performance.

To address these challenges, we propose a dynamic array partitioning approach that optimizes the configuration of the available antennas for either dual-functional signal transmission or echo-signal reception, thereby effectively balancing the demands of both communication and sensing functionalities. The array partitioning configuration is represented by the vector $\mathbf{a}\triangleq [a_1,a_2,\ldots,a_N]^T\in\{0,1\}^N$, where $a_n=1$ indicates that the $n$-th antenna operates as a transmit antenna and $a_n=0$ represents a receive antenna.

The transmitted dual-functional signal at time slot $l$ is
\begin{equation}
    \mathbf{x}[l] = \mathbf{A}\mathbf{W}_\text{c}\mathbf{s}_\text{c}[l] + \mathbf{A}\mathbf{W}_\text{r}\mathbf{s}_\text{r}[l] = \mathbf{A}\mathbf{W}\mathbf{s}[l],
\end{equation}  
where $\mathbf{A}\triangleq \text{diag}\{\mathbf{a}\}$ is the array partitioning matrix, $\mathbf{W} \triangleq [\mathbf{W}_\text{c}~\mathbf{W}_\text{r}]\in\mathbb{C}^{N\times(N+K)}$ is the transmit beamforming matrix, $\mathbf{W}_\text{c}\in\mathbb{C}^{N\times K}$ and $\mathbf{W}_\text{r}\in\mathbb{C}^{N\times N}$ represent the beamformers for the communication symbols $\mathbf{s}_\text{c}\in\mathbb{C}^{K}$ and the radar probing signals $\mathbf{s}_\text{r}\in\mathbb{C}^{N}$, respectively. We assume that $\mathbb{E}\{\mathbf{ss}^H\} = \mathbf{I}_{N+K}$. 
The received signal at the $k$-th user is written as 
\begin{equation}
    y_k[l] = \mathbf{h}_k^T\mathbf{AWs}[l] + n_k[l],
\end{equation}
where $\mathbf{h}_k\in\mathbb{C}^N$ denotes the channel between the BS and the $k$-th user and $n_k[l]\sim\mathcal{CN}(0,\sigma_k^2)$ denotes additive white Gaussian noise (AWGN). The SINR for the $k$-th communication user is given by
\be
\text{SINR}_{k} = \frac{|\mathbf{h}_{k}^T\mathbf{A}\mathbf{w}_{k}|^2}
{\sum_{j\neq k}^{K+N}|\mathbf{h}_{k}^T\mathbf{A}\mathbf{w}_{j}|^2+\sigma_k^2}.
\ee

For sensing, we assume $T \ge 1$ targets are located far enough from the BS to be modeled as point reflectors. The received radar signal at the BS is thus composed of echos from the targets, self-interference between the transmit and receive antennas, and additional noise and interference:  
\be\label{eq:received echo signal}\begin{aligned}
\mathbf{y}_\text{r}[l]  &= \sum_{t=1}^T\alpha_t(\mathbf{I}_N-\mathbf{A})\mathbf{h}_t\mathbf{h}_t^T\mathbf{A}\mathbf{Ws}[l]\\
&\qquad +  \mathbf{H}_\text{SI}\mathbf{Ws}[l+\tau] + (\mathbf{I}_N-\mathbf{A})\mathbf{n}_\text{r}[l],
\end{aligned}\ee
where $\mathbf{H}_\text{SI}\in\mathbb{C}^{N\times N}$ is the self-interference channel, $\tau$ represents the round-trip delay of the target echoes, $\alpha_t\in\mathbb{C}$ is the target radar cross section (RCS), $\mathbf{h}_t\in\mathbb{C}^N$ is the line-of-sight (LoS) channel to the $t$-th target at direction $\theta_t$, i.e., $\mathbf{h}_t \triangleq \beta_t[e^{-\jmath\frac{1-N}{2}\pi\sin\theta_t},e^{-\jmath\frac{3-N}{2}\pi\sin\theta_t},~\ldots,~e^{-\jmath\frac{N-1}{2}\pi\sin\theta_t}]^T$, $\beta_t$ accounts for distance-dependent path loss, and $\mathbf{n}_\text{r}\sim\mathcal{CN}(\mathbf{0},\sigma_\text{r}^2\mathbf{I}_N)$ is AWGN. The radar echoes collected over $L$ time slots can be expressed as
\be\label{eq:Yr}
\mathbf{Y}_\text{r} = \sum_{t=1}^T\alpha_t{\mathbf{H}}_t\mathbf{WS} + \mathbf{H}_\text{s}\mathbf{WSJ}_\tau + (\mathbf{I}_N-\mathbf{A})\mathbf{N}_\text{r},
\ee
where $\mathbf{S}\triangleq [\mathbf{s}[1],\ldots,\mathbf{s}[L]]$, $\mathbf{N}_\text{r}\triangleq [\mathbf{n}_\text{r}[1],\ldots,\mathbf{n}_\text{r}[L]]$, the equivalent channel for the target echoes is  
\be\label{eq:channel Ht}
{\mathbf{H}}_t \triangleq (\mathbf{I}_N-\mathbf{A})\mathbf{h}_t\mathbf{h}_t^T\mathbf{A},
\ee
and the equivalent self-interference channel is defined as
\begin{equation}
\mathbf{H}_\text{s} \triangleq (\mathbf{I}_N-\mathbf{A})\mathbf{H}_\text{SI}\mathbf{A}.
\end{equation}
The time-shift matrix $\mathbf{J}_\tau$ represents the round-trip delay of the target echoes.
We vectorize the echo signal matrix, $\widetilde{\mathbf{y}}_\text{r}=\text{vec}\{\mathbf{Y}_\text{r}\}$, yielding
\begin{equation}\label{eq: yr1}
\widetilde{\mathbf{y}}_\text{r} = 
\underbrace{\sum_{t=1}^T \alpha_t \text{vec}\{{\mathbf{H}}_t \mathbf{WS}\}}_{\bm{\eta} \text{ (signal component)}} + 
\underbrace{\text{vec}\{\mathbf{H}_\text{s} \mathbf{WSJ}_\tau + (\mathbf{I}_N\!-\! \mathbf{A}) \mathbf{N}_\text{r}\}}_{\mathbf{n} \text{ (noise component)}},
\end{equation}
which we assume to follow a complex Gaussian distribution: $\widetilde{\mathbf{y}}_\text{r}\sim\mathcal{CN}(\bm{\eta},\widetilde{\mathbf{R}}_\text{n})$ with covariance matrix $\widetilde{\mathbf{R}}_\text{n} = \mathbf{I}_L\otimes \mathbf{R}_\text{n}$, where $\mathbf{R}_\text{n}$ is the covariance of the self-interference plus noise: 
\be\label{eq:Rn}
\mathbf{R}_\text{n} = (\mathbf{I}_N-\mathbf{A})\big[\sigma_\text{r}^2\mathbf{I}_N + \mathbf{H}_\text{SI}\mathbf{AWW}^H\mathbf{A}\mathbf{H}_\text{SI}^H\big](\mathbf{I}_N-\mathbf{A}).\ee

Given $\widetilde{\mathbf{y}}_\text{r}$, the BS aims to estimate the target parameters 
\begin{subequations}\begin{align}
\bm{\theta}&\triangleq [\theta_1,\theta_2,\ldots,\theta_T]^T,\\ \bm{\alpha}&\triangleq[\alpha_1,\alpha_2,\ldots,\alpha_T]^T,\\
\bm{\xi}&\triangleq[\bm{\theta}^T,\Re\{\bm{\alpha}^T\},\Im\{\bm{\alpha}^T\}]^T.
\end{align}\end{subequations}
The likelihood function for parameter estimation is given by
\begin{equation}\label{eq:LLF}
p(\widetilde{\mathbf{y}}_{\text{r}} | \bm{\theta}, \bm{\alpha}) = \frac{\exp\big[-(\widetilde{\mathbf{y}}_{\text{r}} - \bm{\eta})^H \widetilde{\mathbf{R}}_\text{n}^{-1} (\widetilde{\mathbf{y}}_{\text{r}} - \bm{\eta})\big]}{\pi^{NL} |\widetilde{\mathbf{R}}_\text{n}|}.
\end{equation}

Incorporating prior knowledge about the target parameters can greatly improve estimation accuracy. In tracking scenarios, the BS can leverage historical observations to obtain statistical priors. In our proposed approach, we assume the following prior distributions of the target DOAs and RCS:
\begin{equation}
    \bm{\bm{\theta}} \sim \mathcal{N}(\bm{\mu}_\theta, \bm{\Sigma}_\theta), \quad \bm{\alpha} \sim \mathcal{CN}(\bm{\mu}_\alpha, \bm{\Sigma}_\alpha),
\end{equation}
with the corresponding prior probabilities 
\begin{subequations}\label{eq:prior prob}\begin{align}
p(\bm{\theta}) &= \frac{\exp \left[ -\frac{1}{2} (\bm{\theta} - \bm{\mu}_\theta)^T \bm{\Sigma}_\theta^{-1} (\bm{\theta} - \bm{\mu}_\theta) \right]}{(2\pi)^{T/2}|\bm{\Sigma}_\theta|^{1/2}},\\
p(\bm{\alpha}) &= \frac{\exp \left[ -(\bm{\alpha} - \bm{\mu}_\alpha)^H \bm{\Sigma}_\alpha^{-1} (\bm{\alpha} - \bm{\mu}_\alpha) \right]}{\pi^T|\bm{\Sigma}_\alpha|}.
\end{align}\end{subequations}
Combining the prior information with the likelihood, the posterior probability for estimating $\bm{\theta}$ and $\bm{\alpha}$ is given by
\begin{equation}
p(\bm{\theta}, \bm{\alpha} | \widetilde{\mathbf{y}}_r) \propto p(\widetilde{\mathbf{y}}_r | \bm{\theta}, \bm{\alpha}) p(\bm{\theta}) p(\bm{\alpha}).
\label{eq:posterior}
\end{equation}
Ignoring constants independent of $(\bm{\theta},\bm{\alpha})$, the negative log-posterior can be expressed as 
\begin{equation}\begin{aligned}\label{eq:J}
\mathcal{L}(\bm{\theta},\bm{\alpha}) &= (\widetilde{\mathbf{y}}_\text{r} \!-\! \bm{\eta})^H \widetilde{\mathbf{R}}^{-1}_\text{n} (\widetilde{\mathbf{y}}_\text{r} \!-\! \bm{\eta}) + \frac{1}{2}(\bm{\theta} \!-\! \bm{\mu}_\theta)^T \bm{\Sigma}_{\theta}^{-1} (\bm{\theta} \!-\! \bm{\mu}_{\theta}) \\
&\quad + (\bm{\alpha} - \bm{\mu}_\alpha)^T \bm{\Sigma}_{\alpha}^{-1} (\bm{\alpha} - \bm{\mu}_\alpha).
\end{aligned}\end{equation}

Estimates of the target parameters $\bm{\theta}$ and $\bm{\alpha}$ can be obtained by maximizing the posterior probability, which is equivalent to minimizing the negative log-posterior $\mathcal{L}(\bm{\theta},\bm{\alpha})$. The sensing performance is typically evaluated by the mean squared error (MSE) of the parameter estimates. However, due to the complexity of the sensing scenario, deriving closed-form expressions for the estimates or their associated errors is often intractable. As an alternative, a widely adopted approach is to evaluate the theoretical lower bound on the estimation errors using the Bayesian Cramér-Rao bound (BCRB). The BCRB provides a fundamental limit on the achievable estimation accuracy by leveraging the Fisher information matrix (FIM), which quantifies the overall information available for parameter estimation. The Bayesian FIM is defined as  \cite{Kay-book1998}:
\begin{equation}
    \mathbf{F}_\text{B} = -\mathbb{E}\Big\{\frac{\partial^2\ln p(\bm{\theta},\bm{\alpha}|\widetilde{\mathbf{y}}_\text{r})}{\partial\bm{\xi}^2}\Big\}.
\end{equation}
According to \eqref{eq:posterior} and \eqref{eq:J}, $\mathbf{F}_\text{B}$ consists of two components:
\begin{equation}
    \mathbf{F}_\text{B} = \mathbf{F}_\text{L} + \mathbf{F}_\text{P},
\end{equation}
where $\mathbf{F}_\text{L}$ is the likelihood-based FIM that captures the information provided by the observations, while $\mathbf{F}_\text{P}$ reflects the information contained in the prior distributions. 
Given the prior probabilities in \eqref{eq:prior prob} and assuming independence between $\bm{\theta}$ and $\bm{\alpha}$, the prior FIM is expressed as 
\begin{equation}\label{eq:Fp partition}
    \mathbf{F}_{\text{P}} = 
    \begin{bmatrix}
    \mathbf{F}_\theta &  \\
    & \mathbf{F}_\alpha
    \end{bmatrix},
\end{equation}
where 
$\mathbf{F}_\theta = \bm{\Sigma}_{\theta}^{-1}$ and $\mathbf{F}_\alpha = \begin{bmatrix}
    2\Re\{\bm{\Sigma}_\alpha^{-1}\} & 2\Im\{\bm{\Sigma}_\alpha^{-1}\} \\
    -2\Im\{\bm{\Sigma}_\alpha^{-1}\} & 2\Im\{\bm{\Sigma}_\alpha^{-1}\}
\end{bmatrix}$.

The likelihood-based FIM $\mathbf{F}_\text{L}\in\mathbb{R}^{3T\times 3T}$ is defined by 
\begin{equation}\label{eq:FL ij}
\mathbf{F}_\text{L}(i,j) = 2\Re\Big\{\frac{\partial\bm{\eta}^H}{\partial\xi_i}\widetilde{\mathbf{R}}_\text{n}^{-1}\frac{\partial\bm{\eta}}{\partial\xi_j}\Big\},
\end{equation}
with partial derivatives  
\begin{subequations}\label{eq:eta2xi}\begin{align}
\frac{\partial\bm{\eta}}{\partial\theta_t} &=\alpha_t \text{vec}\{\dot{{\mathbf{H}}}_t\mathbf{WS}\}, \\
\frac{\partial\bm{\eta}}{\partial\Re\{\alpha_t\}} &= \text{vec}\{{\mathbf{H}}_t\mathbf{WS}\}, \\
\frac{\partial\bm{\eta}}{\partial\Im\{\alpha_t\}} &= \jmath\text{vec}\{{\mathbf{H}}_t\mathbf{WS}\},
\end{align}\end{subequations}
and $\dot{\mathbf{H}}_t = \partial \mathbf{H}_t / \partial \theta_t$. 
Recalling \eqref{eq:channel Ht}, we can write
\begin{subequations}\label{eq:Ht Htd}
\begin{align}
\mathbf{H}_t&= \text{diag}\{\mathbf{h}_t\}\mathbf{ba}^T\text{diag}\{\mathbf{h}_t\},\\
\dot{\mathbf{H}}_t &= (\mathbf{I}_N-\mathbf{A})\dot{\mathbf{h}}_t\mathbf{h}_t^T\mathbf{A} +  (\mathbf{I}_N-\mathbf{A})\mathbf{h}_t\dot{\mathbf{h}}_t^T\mathbf{A},\\
& = \text{diag}\{\dot{\mathbf{h}}_t\}\mathbf{ba}^T\text{diag}\{\mathbf{h}_t\} + \text{diag}\{\mathbf{h}_t\}\mathbf{ba}^T\text{diag}\{\dot{\mathbf{h}}_t\},
\end{align}\end{subequations}
where $\mathbf{b}\triangleq \mathbf{1}-\mathbf{a}$, and $\dot{\mathbf{h}}_\text{t}\triangleq\partial\mathbf{h}_\text{t}/\partial\theta_\text{t}= -\jmath\pi\cos\theta_\text{t}\mathbf{Q}\mathbf{h}_\text{t}$ with $\mathbf{Q}\triangleq\text{diag}\{\mathbf{q}\}$ and $\mathbf{q}\triangleq [\frac{1-N}{2},\frac{3-N}{2},\ldots,\frac{N-1}{2}]^T$. 
Note that both $\mathbf{H}_t$ and $\dot{\mathbf{H}}_t$ depend on the array partitioning vector $\mathbf{a}$. Using the definitions in \eqref{eq:FL ij} and \eqref{eq:eta2xi}, the elements of $\mathbf{F}_\text{L}$ are shown in \eqref{eq:FIM} on the next page, where $\mathbf{R}_\text{w} = \mathbf{WW}^H$. 
For convenience in further derivations, we partition $\mathbf{F}_\text{L}$ as
\be\label{eq:FL partition}
\mathbf{F}_\text{L} = \begin{bmatrix}
\mathbf{F}_{\bm{\theta}\bm{\theta}^T}  & \mathbf{F}_{\bm{\theta}\widetilde{\bm{\alpha}}^T} \\
\mathbf{F}_{\bm{\theta}\widetilde{\bm{\alpha}}^T}^T & \mathbf{F}_{\widetilde{\bm{\alpha}}\widetilde{\bm{\alpha}}^T} 
\end{bmatrix}, \ee
where $\mathbf{F}_{\bm{\theta}\bm{\theta}^T}\in\mathbb{C}^{T\times T}$, $\mathbf{F}_{\bm{\theta}\widetilde{\bm{\alpha}}^T}\in\mathbb{C}^{T\times 2T}$, and $\mathbf{F}_{\widetilde{\bm{\alpha}}\widetilde{\bm{\alpha}}^T}\in\mathbb{C}^{2T\times 2T}$. 

Combining the prior and likelihood-based FIMs, the Bayesian FIM is expressed as 
\begin{equation}
\mathbf{F}_\text{B} =
\begin{bmatrix}
\mathbf{F}_{\bm{\theta}\bm{\theta}^T} + \mathbf{F}_\theta & \mathbf{F}_{\bm{\theta}\widetilde{\bm{\alpha}}^T} \\
\mathbf{F}_{\bm{\theta}\widetilde{\bm{\alpha}}^T}^T & \mathbf{F}_{\widetilde{\bm{\alpha}}\widetilde{\bm{\alpha}}^T} + \mathbf{F}_\alpha
\end{bmatrix},
\end{equation}
and the BCRB is obtained by inverting $\mathbf{F}_\text{B}$. We will focus on the diagonal elements of the BCRB, which represent lower bounds on the estimation error variance for each parameter in $\bm{\xi}$.
Our focus will be on DOA estimation performance, so we are interested in reducing the trace of the BCRB corresponding only to $\bm{\theta}$:
\begin{equation}
\text{BCRB}_{\bm{\theta}} = \text{Tr}\big\{\big[\mathbf{F}_{\bm{\theta}\bm{\theta}^T}+\mathbf{F}_\theta-\mathbf{F}_{\bm{\theta}\widetilde{\bm{\alpha}}^T}^T(\mathbf{F}_{\widetilde{\bm{\alpha}}\widetilde{\bm{\alpha}}^T}+\mathbf{F}_\alpha)^{-1}\mathbf{F}_{\bm{\theta}\widetilde{\bm{\alpha}}^T}]^{-1}\big\}.
\end{equation}
More generally, our estimation criterion will involve minimizing the following weighted trace: 
\begin{equation}\label{eq:objective}
\text{Tr}\big\{\bm{\Lambda}\big[\mathbf{F}_{\bm{\theta}\bm{\theta}^T}+\mathbf{F}_\theta-\mathbf{F}_{\bm{\theta}\widetilde{\bm{\alpha}}^T}^T(\mathbf{F}_{\widetilde{\bm{\alpha}}\widetilde{\bm{\alpha}}^T}+\mathbf{F}_\alpha)^{-1}\mathbf{F}_{\bm{\theta}\widetilde{\bm{\alpha}}^T}]^{-1}\big\},    
\end{equation}
with weighting matrix $\bm{\Lambda} = \text{diag}\{\lambda_1^2,\lambda_2^2,\ldots,\lambda_T^2\}$.

\newcounter{TempEqCnt}
\setcounter{TempEqCnt}{\value{equation}}
\setcounter{equation}{25}
In the next section, we will study minimization of the BCRB for DOA estimation, subject to constraints on communication SINR, transmit power, and the array partitioning, as formulated in the following optimization problem: 
\begin{subequations}\label{eq:original problem}\begin{align}
&\underset{\mathbf{a},\mathbf{W}}\min~~\text{Tr}\big\{\bm{\Lambda}\big[\mathbf{F}_{\bm{\theta}\bm{\theta}^T}\!+\!\mathbf{F}_\theta-\mathbf{F}_{\bm{\theta}\widetilde{\bm{\alpha}}^T}^T(\mathbf{F}_{\widetilde{\bm{\alpha}}\widetilde{\bm{\alpha}}^T}\!+\!\mathbf{F}_\alpha)^{-1}\mathbf{F}_{\bm{\theta}\widetilde{\bm{\alpha}}^T}]^{-1}\big\}\label{eq:original problem obj}\\
&~~\text{s.t.}~~\frac{|\mathbf{h}_{k}^T\mathbf{A}\mathbf{w}_{k}|^2}
{\sum_{j\neq k}^{N+K}|\mathbf{h}_{k}^T\mathbf{A}\mathbf{w}_{j}|^2+\sigma_k^2}\geq \Gamma_k,~\forall k, \label{eq:original c1}\\
&\qquad~~\|\mathbf{AW}\|_F^2 \leq P,\label{eq:original c2}\\
&\qquad~~K\leq\mathbf{1}^T\mathbf{a}\leq N-T,\label{eq:original c3}\\
&\qquad~~a_n\in\{0,1\},~\forall n,\label{eq:original c4}
\end{align}
\end{subequations}
where $\Gamma_k$ is the communication SINR threshold for the $k$-th user, $P$ is the transmit power budget, and \eqref{eq:original c3} is imposed to guarantee $K$ different data streams and the desired sensing capability. This optimization problem is inherently complex and highly non-convex, primarily due to the fractional and quadratic terms in the objective and constraints, as well as the binary integer constraints on the array partitioning. In the next section, we introduce an efficient alternating optimization algorithm that decomposes the problem into several manageable sub-problems and solves them iteratively. Subsequently, in Section IV, we extend the joint array partitioning and transmit beamforming design to address the extended target scenario.

\setcounter{equation}{26}
\begin{figure*}
\begin{subequations}\label{eq:FIM}\begin{align}
\mathbf{F}_\text{L}(i,j) & =2\Re\big\{\alpha_i^*\alpha_j\text{vec}^H\{\dot{\mathbf{H}}_i\mathbf{WS}\}(\mathbf{I}_L\otimes\mathbf{R}_\text{n})^{-1}\text{vec}\{\dot{\mathbf{H}}_j\mathbf{WS}\}\big\}  \\
& = 2L\Re\big\{\alpha_i^*\alpha_j\text{Tr}\{\mathbf{R}_\text{n}^{-1}\dot{\mathbf{H}}_j
\mathbf{R}_\text{w}\dot{\mathbf{H}}_i^H\big\} \\
& = 2L\Re\big\{\alpha_i^*\alpha_j\big[\mathbf{a}^T\text{diag}\{{\mathbf{h}}_j\}\mathbf{R}_\text{w}\text{diag}\{{\mathbf{h}}_i^H\}\mathbf{a}\mathbf{b}^T\text{diag}\{\dot{\mathbf{h}}_i^H\}\mathbf{R}^{-1}_\text{n}\text{diag}\{\dot{\mathbf{h}}_j\}\mathbf{b} \non \\
&\qquad\qquad\qquad~ + \mathbf{a}^T\text{diag}\{\dot{\mathbf{h}}_j\}\mathbf{R}_\text{w}\text{diag}\{\mathbf{h}_i^H\}\mathbf{a}\mathbf{b}^T\text{diag}\{\dot{\mathbf{h}}_i^H\}\mathbf{R}^{-1}_\text{n}\text{diag}\{{\mathbf{h}}_j\}\mathbf{b} \non\\
&\qquad\qquad\qquad~ + \mathbf{a}^T\text{diag}\{\mathbf{h}_j\}\mathbf{R}_\text{w}\text{diag}\{\dot{\mathbf{h}}_i^H\}\mathbf{a}\mathbf{b}^T\text{diag}\{\mathbf{h}_i^H\}\mathbf{R}^{-1}_\text{n}\text{diag}\{\dot{\mathbf{h}}_j\}\mathbf{b} \non\\
&\qquad\qquad\qquad~ + 
\mathbf{a}^T\text{diag}\{\dot{\mathbf{h}}_j\}\mathbf{R}_\text{w}\text{diag}\{\dot{\mathbf{h}}_i^H\}\mathbf{a}\mathbf{b}^T\text{diag}\{\mathbf{h}_i^H\}\mathbf{R}^{-1}_\text{n}\text{diag}\{{\mathbf{h}}_j\}\mathbf{b}\big]\big\},~\forall i,~j = 1,2,\ldots,T,\\
\mathbf{F}_\text{L}(i,T+j) & = 2L\Re\big\{\alpha_i^*\text{Tr}\{\mathbf{R}_\text{n}^{-1}\mathbf{H}_j
\mathbf{R}_\text{w}\dot{\mathbf{H}}_i^H\big\}\\
&= 2L\Re\big\{\alpha_i^*\big[\mathbf{a}^T\text{diag}\{{\mathbf{h}}_j\}\mathbf{R}_\text{w}\text{diag}\{{\mathbf{h}}_i^H\}\mathbf{a}\mathbf{b}^T\text{diag}\{\dot{\mathbf{h}}_i^H\}\mathbf{R}_\text{n}^{-1}\text{diag}\{{\mathbf{h}}_j\}\mathbf{b} \non\\
&\qquad\qquad\quad~ + \mathbf{a}^T\text{diag}\{{\mathbf{h}}_j\}\mathbf{R}_\text{w}\text{diag}\{\dot{\mathbf{h}}_i^H\}\mathbf{a}\mathbf{b}^T\text{diag}\{{\mathbf{h}}_i^H\}\mathbf{R}_\text{n}^{-1}\text{diag}\{{\mathbf{h}}_j\}\mathbf{b}\big]\big\},~\forall i,~j = 1,2,\ldots,T,\\
\mathbf{F}_\text{L}(i,2T+j) & = 2L\Re\big\{\jmath\alpha_i^*\text{Tr}\{\mathbf{R}_\text{n}^{-1}\mathbf{H}_j
\mathbf{R}_\text{w}\dot{\mathbf{H}}_i^H\big\},~\forall i,~j = 1,2,\ldots,T,\\
\mathbf{F}_\text{L}(T+i,T+j) & = \mathbf{F}_\text{L}(2T+i,2T+j) = 2L\Re\big\{\text{Tr}\{\mathbf{R}_\text{n}^{-1}\mathbf{H}_j
\mathbf{R}_\text{w}\mathbf{H}_i^H\big\}\\
& = 2L\Re\big\{\mathbf{a}^T\text{diag}\{{\mathbf{h}}_j\}\mathbf{R}_\text{w}\text{diag}\{{\mathbf{h}}_i^H\}\mathbf{a}\mathbf{b}^T\text{diag}\{{\mathbf{h}}_i^H\}\mathbf{R}_\text{n}^{-1}\text{diag}\{{\mathbf{h}}_j\}\mathbf{b}\big\},~\forall i,~j = 1,2,\ldots,T,\\
\mathbf{F}_\text{L}(T+i,2T+j) & = 2L\Re\big\{\jmath\text{Tr}\{\mathbf{R}_\text{n}^{-1}\mathbf{H}_j
\mathbf{R}_\text{w}\mathbf{H}_i^H\big\},\\
\mathbf{F}_\text{L}(2T+i,T+j) & = 2L\Re\big\{-\jmath\text{Tr}\{\mathbf{R}_\text{n}^{-1}\mathbf{H}_j
\mathbf{R}_\text{w}\mathbf{H}_i^H\big\},~\forall i,~j = 1,2,\ldots,T.
\end{align}\end{subequations}
\rule[-0pt]{18.5 cm}{0.05em}
\end{figure*}

\section{Joint Array Partitioning and Transmit Beamforming Design For Point-Like Targets}
In this section, we propose an alternating optimization algorithm to solve the joint array partitioning and transmit beamforming design problem for the multiple point-like-target scenario. The Schur complement is first used to reformulate the objective function into a more tractable form. Subsequently, the binary integer constraint is relaxed by introducing a smooth penalty term along with convex constraints. To further facilitate the optimization process, the ADMM framework is employed to decompose the problem into a series of tractable subproblems that are then iteratively solved using block coordinate descent. The detailed algorithmic steps are provided in the following subsections.

\subsection{Schur Complement Transformation}

To address the complex objective function, we introduce an  auxiliary variable $\mathbf{U}\in\mathbb{S}_T^{+}$ and leverage the Schur complement to convert the optimization problem to 
\begin{subequations}\begin{align}
&\underset{\mathbf{a},\mathbf{W},\mathbf{U}\in\mathbb{S}_T^{+}}\min
~~\text{Tr}\{\mathbf{U}^{-1}\} \\
&\qquad\text{s.t.}\quad \left[\begin{matrix} \mathbf{F}_{\bm{\theta}\bm{\theta}^T}+\mathbf{F}_\theta-\mathbf{U}\bm{\Lambda}& \mathbf{F}_{\bm{\theta}\widetilde{\bm{\alpha}}^T}\\ \mathbf{F}_{\bm{\theta}\widetilde{\bm{\alpha}}^T}^T& \mathbf{F}_{\widetilde{\bm{\alpha}}\widetilde{\bm{\alpha}}^T+\mathbf{F}_\alpha}\end{matrix}\right]\succeq\mathbf{0},\label{eq:Schur comple}\\
&\qquad\qquad~\eqref{eq:original c1}-\eqref{eq:original c4}.
\end{align}\end{subequations}

\subsection{Binary Integer Constraint}
Since the binary integer constraint \eqref{eq:original c4} introduces combinatorial complexity that makes the problem challenging to solve directly, we approximate this constraint by reformulating it as a smooth penalty term in the objective function accompanied by a box constraint:
\begin{subequations}\begin{align}
&\underset{\mathbf{a},\mathbf{W},\mathbf{U}\in\mathbb{S}_T^{+}}\min~
\text{Tr}\{\mathbf{U}^{-1}\}+\rho_1\mathbf{a}^T(\mathbf{1}-\mathbf{a}) \\
&\qquad\text{s.t.}\quad~ 0\leq a_n\leq 1,~\forall n,\\
&\qquad\qquad~~\eqref{eq:original c1}-\eqref{eq:original c3},~ \eqref{eq:Schur comple},
\end{align}\end{subequations}
where $\rho_1>0$ is a preset penalty parameter that regulates the extent to which the binary integer constraint is enforced.

\subsection{ADMM-Based Transformation}
We observe that both the objective and the constraints involve quadratic terms with respect to $\mathbf{b}$, where $\mathbf{b} \triangleq \mathbf{1}-\mathbf{a}$. To simplify the problem, it is natural to introduce $\mathbf{b} = \mathbf{1}-\mathbf{a}$ as an auxiliary variable. Doing so and employing the ADMM framework \cite{ADMM}, we express the corresponding augmented Lagrangian formulation of the problem as
\begin{subequations}\label{eq:ALp}
\begin{align}
&\underset{\mathbf{a},\mathbf{W},\mathbf{U}\in\mathbb{S}_T^{+},\mathbf{b}}\min
\text{Tr}\{\mathbf{U}^{-1}\}+\rho_1\mathbf{a}^T(\mathbf{1}-\mathbf{a}) +\rho_2\big\|\mathbf{b}-\mathbf{1}+\mathbf{a}+\frac{\bm{\mu}}{\rho_2}\big\|^2 \label{eq:obj1} \\
&\qquad~\text{s.t.}\quad~ 0 \leq a_n, b_n \leq 1,~\forall n,\\
&\qquad\qquad\quad T\leq\mathbf{1}^T\mathbf{b}\leq N-K,\\
&\qquad\qquad\quad \eqref{eq:original c1}-\eqref{eq:original c3},~ \eqref{eq:Schur comple},
\end{align}\end{subequations}
where $\rho_2>0$ is a penalty parameter and $\bm{\mu}\in\mathbb{R}^N$ is the dual variable. 
Then, we employ the block coordinate descent method to solve this multivariate optimization problem. In each iteration, the dual variable is updated by 
\be\label{eq:update mu}
\bm{\mu} := \bm{\mu} + \rho_2(\mathbf{b}-\mathbf{1} + \mathbf{a}).
\ee
The update rules for other variables are presented in detail in the following subsections. 

\subsection{Update $\mathbf{W}$}
Given solutions for the other variables, the sub-problem for finding $\mathbf{W}$ is formulated as 
\begin{equation}\label{eq:W prob}\begin{aligned}
&\underset{\mathbf{W},\mathbf{U}\in\mathbb{S}_T^{+}}\min~
\text{Tr}\{\mathbf{U}^{-1}\}\\
&\quad~\text{s.t.}\quad~\eqref{eq:original c1},~\eqref{eq:original c2},~\eqref{eq:Schur comple}.
\end{aligned}\end{equation}
Since both the objective and constraints involve quadratic terms with respect to $\mathbf{W}$, we define $\mathbf{R}_\text{w}\triangleq\mathbf{WW}^H$ and $\mathbf{R}_k\triangleq\mathbf{w}_k\mathbf{w}_k^H,~\forall k$. Then using the typical SDR approach \cite{ZLuo-SPM-2010}, we transform problem \eqref{eq:W prob} as 
\begin{subequations}\label{eq:solve for W}\begin{align}
&\underset{\mathbf{U}\in\mathbb{S}_T^{+},\mathbf{R}_\text{w},\mathbf{R}_k,\forall k}\min
~~\text{Tr}\{\mathbf{U}^{-1}\}\\
&\quad\text{s.t.}~\left[\begin{matrix} \mathbf{F}_{\bm{\theta}\bm{\theta}^T}(\mathbf{R}_\text{w})+\mathbf{F}_\theta-\mathbf{U}\bm{\Lambda}& \mathbf{F}_{\bm{\theta}\widetilde{\bm{\alpha}}^T}(\mathbf{R}_\text{w})\\ \mathbf{F}_{\bm{\theta}\widetilde{\bm{\alpha}}^T}^T(\mathbf{R}_\text{w})& \mathbf{F}_{\widetilde{\bm{\alpha}}\widetilde{\bm{\alpha}}^T}(\mathbf{R}_\text{w})+\mathbf{F}_\alpha\end{matrix}\right]\succeq\mathbf{0}, \label{eq:constraint in Rw}\\
&\qquad\quad(1+\Gamma_k^{-1})\mathbf{a}^T\text{diag}\{\mathbf{h}_k\}\mathbf{R}_k\text{diag}\{\mathbf{h}_k^H\}\mathbf{a} \nonumber\\
&\qquad\qquad -\mathbf{a}^T\text{diag}\{\mathbf{h}_k\}\mathbf{R}_\text{w}\text{diag}\{\mathbf{h}_k^H\}\mathbf{a}\geq \sigma_k^2,~\forall k, \\
&\qquad\quad\text{Tr}\{\text{diag}\{\mathbf{a}\}\mathbf{R}_\text{w}\text{diag}\{\mathbf{a}\}\} \leq P,\\
&\qquad\quad\mathbf{R}_\text{w},~\mathbf{R}_k,~\forall k,~\mathbf{R}_\text{w}-\sum\nolimits_{k=1}^K\mathbf{R}_k\in\mathbb{S}_N^+. \end{align}\end{subequations}
In \eqref{eq:constraint in Rw}, the matrices $\mathbf{F}_{\bm{\theta}\bm{\theta}^T}$, $\mathbf{F}_{\bm{\theta}\widetilde{\bm{\alpha}}^T}$ and $\mathbf{F}_{\widetilde{\bm{\alpha}}\widetilde{\bm{\alpha}}^T}$ are linear functions of $\mathbf{R}_\text{w}$ as shown in \eqref{eq:FIM}. 
This is a semi-definite programming (SDP) problem that can be efficiently solved using conventional optimization toolboxes.

After obtaining the solutions $\widetilde{\mathbf{R}}_\text{w}$ and $\widetilde{\mathbf{R}}_k$ to \eqref{eq:solve for W}, we have the optimal solution $\mathbf{R}_\text{w}= \widetilde{\mathbf{R}}_\text{w}$, and we can find the optimal $\mathbf{R}_k$ that satisfy $\text{Rank}\{\mathbf{R}_k\} = 1,~\forall k$, using 
\begin{subequations}\label{eq:update w}\begin{align}
\mathbf{w}_k &= (\mathbf{h}_k^T\mathbf{A}\widetilde{\mathbf{R}}_k\mathbf{A}\mathbf{h}_k^*)^{-1/2}\widetilde{\mathbf{R}}_k\mathbf{A}\mathbf{h}_k^*,~\forall k,\label{eq:initial wk}\\
\mathbf{R}_k &=  \mathbf{w}_k\mathbf{w}_k^H,
\end{align}\end{subequations}
where $\mathbf{w}_k$ is the $k$-th column of the communication beamformer $\mathbf{W}_\text{c}$. 
Then, recalling that $\mathbf{R}_\text{w} = \mathbf{WW}^H=\mathbf{W}_\text{c}\mathbf{W}_\text{c}^H+\mathbf{W}_\text{r}\mathbf{W}_\text{r}^H$, the radar beamformer $\mathbf{W}_\text{r}$ should satisfy
\be\label{eq:update Wr}
\mathbf{W}_\text{r}\mathbf{W}_\text{r}^H = \mathbf{R}_\text{w}-\sum_{k=1}^K\mathbf{R}_k,
\ee
from which $\mathbf{W}_\text{r}$ can be found using either a Cholesky or eigenvalue decomposition.
\vspace{-0.3 cm}
\subsection{Update $\mathbf{a}$}
After obtaining the other variables, the sub-problem for updating $\mathbf{a}$ is formulated as 
\begin{subequations}\label{eq:a prob}\begin{align}
&\underset{\mathbf{U}\in\mathbb{S}_T^{+},\mathbf{a}}\min
~~\text{Tr}\{\mathbf{U}^{-1}\} + \rho_1\mathbf{a}^T(\mathbf{1}-\mathbf{a})+\rho_2\|\mathbf{b}-\mathbf{1}+\mathbf{a}+\bm{\mu}/\rho_2\|^2    \\
&\quad\text{s.t.}~ \left[\begin{matrix} \mathbf{F}_{\bm{\theta}\bm{\theta}^T}(\mathbf{a})+\mathbf{F}_\theta-\mathbf{U}\bm{\Lambda}& \mathbf{F}_{\bm{\theta}\widetilde{\bm{\alpha}}^T}(\mathbf{a})\\ \mathbf{F}_{\bm{\theta}\widetilde{\bm{\alpha}}^T}^T(\mathbf{a})& \mathbf{F}_{\widetilde{\bm{\alpha}}\widetilde{\bm{\alpha}}^T}(\mathbf{a})+\mathbf{F}_\alpha\end{matrix}\right]\succeq\mathbf{0},\\
&\qquad\quad (1+\Gamma_k^{-1})\mathbf{a}^T\text{diag}\{\mathbf{h}_k\}\mathbf{w}_k\mathbf{w}_k^H\text{diag}\{\mathbf{h}_k^H\}\mathbf{a} \nonumber\\
&\qquad\quad~-\mathbf{a}^T\text{diag}\{\mathbf{h}_k\}\mathbf{WW}^H\text{diag}\{\mathbf{h}_k^H\}\mathbf{a}\geq \sigma_k^2,~\forall k, \\
&\qquad\quad\text{Tr}\{\text{diag}\{\mathbf{a}\}\mathbf{WW}^H\text{diag}\{\mathbf{a}\}\} \leq P,\\
&\qquad\quad K^2\leq\mathbf{a}^T\mathbf{1}_N\mathbf{a}\leq (N-T)^2,\\
&\qquad\quad 0\leq a_n^2\leq 1,~\forall n.
\end{align}\end{subequations}
This problem involves both quadratic and linear functions of $\mathbf{a}$ along with constraints that include fractional expressions and semidefinite conditions. To address these difficulties, we define the variable $\widetilde{\mathbf{a}}\triangleq [\mathbf{a}^T~1]^T$, and the primary variable 
$\widetilde{\mathbf{A}}\triangleq  \widetilde{\mathbf{a}}\widetilde{\mathbf{a}}^T 
 = \left[\begin{matrix}\widetilde{\mathbf{A}}_1 &\mathbf{a}\\
\mathbf{a}^T&1\end{matrix}\right]$ where  $\widetilde{\mathbf{A}}_1 \triangleq \mathbf{aa}^T$.
After some matrix manipulations and using the SDR approach, problem \eqref{eq:a prob} can be transformed as 
\begin{subequations}\label{eq:solve for A}\begin{align}
&\underset{\mathbf{U}\in\mathbb{S}_T^{+},\widetilde{\mathbf{A}}}\min
~~\text{Tr}\{\mathbf{U}^{-1}\} + \text{Tr}\{\widetilde{\mathbf{A}}\mathbf{E}_b\}      \\
&\quad\text{s.t.}\quad \left[\begin{matrix} \mathbf{F}_{\bm{\theta}\bm{\theta}^T}(\widetilde{\mathbf{A}}_1)+\mathbf{F}_\theta-\mathbf{U}\bm{\Lambda}& \mathbf{F}_{\bm{\theta}\widetilde{\bm{\alpha}}^T}(\widetilde{\mathbf{A}}_1)\\ \mathbf{F}_{\bm{\theta}\widetilde{\bm{\alpha}}^T}^T(\widetilde{\mathbf{A}}_1)& \mathbf{F}_{\widetilde{\bm{\alpha}}\widetilde{\bm{\alpha}}^T}(\widetilde{\mathbf{A}}_1)+\mathbf{F}_\alpha\end{matrix}\right]\succeq\mathbf{0},\label{eq:cons4a}\\
&\qquad\quad~\text{Tr}\{\mathbf{D}_k\widetilde{\mathbf{A}}_1\}\geq \sigma_k^2,~\forall k, \\
&\qquad\quad~\sum_{j=1}^{K+N}\text{Tr}\big\{\text{diag}\{|\mathbf{w}_j|^2\}\widetilde{\mathbf{A}}_1\big\} \leq P,\\
&\qquad\quad~K^2\leq\text{Tr}\{\widetilde{\mathbf{A}}_1\mathbf{1}\}\leq (N-T)^2,\\
&\qquad\quad~0\leq \widetilde{\mathbf{A}}(n,n)\leq 1,~\forall n,\\
&\qquad\quad~\widetilde{\mathbf{A}}(N+1,N+1) = 1 ,
\end{align}\end{subequations}
where the matrices $\mathbf{F}_{\bm{\theta}\bm{\theta}^T}$, $\mathbf{F}_{\bm{\theta}\widetilde{\bm{\alpha}}^T}$ and $\mathbf{F}_{\widetilde{\bm{\alpha}}\widetilde{\bm{\alpha}}^T}$ can be easily expressed as linear functions with respect to $\widetilde{\mathbf{A}}_1$, i.e., $\mathbf{aa}^T$, according to the expressions in \eqref{eq:FIM}, and we define
\begin{equation}\begin{aligned}
\mathbf{E}_b&= \left[\begin{matrix}(\rho_2-\rho_1)\mathbf{I}_N &\mathbf{c}_b\\
\mathbf{c}_b^T&0\end{matrix}\right],\\
\mathbf{c}_b &= 0.5\rho_1\mathbf{1} + \rho_2(\mathbf{b}-\mathbf{1}+\bm{\mu}/\rho_2),\\
\mathbf{D}_k & =  (1+\Gamma_k^{-1})\text{diag}\{\mathbf{h}_k\}\mathbf{w}_k\mathbf{w}_k^H\text{diag}\{\mathbf{h}_k^H\} \\ & \qquad -\text{diag}\{\mathbf{h}_k\}\mathbf{WW}^H\text{diag}\{\mathbf{h}_k^H\}.
\end{aligned}\end{equation}
After obtaining $\widetilde{\mathbf{A}}_1$ by solving problem \eqref{eq:solve for A}, we can construct the optimal solution as
\be\label{eq:update a}
\mathbf{a} = (\mathbf{1}^T\widetilde{\mathbf{A}}_1\mathbf{1})^{-1/2}\widetilde{\mathbf{A}}_1\mathbf{1}
\ee
if the rank-one constraint is satisfied; otherwise Gaussian randomization is necessary to recover an approximate solution.
\vspace{-0.3 cm}
\subsection{Update $\mathbf{b}$} 
Fixing the other variables, the update procedure for $\mathbf{b}$ is similar to that for $\mathbf{a}$. In particular, defining 
$\widetilde{\mathbf{b}}\triangleq [\mathbf{b}^T~1]^T$ and 
$\widetilde{\mathbf{B}}\triangleq  \widetilde{\mathbf{b}}\widetilde{\mathbf{b}}^T 
 = \left[\begin{matrix}\widetilde{\mathbf{B}}_1 &\mathbf{b}\\
\mathbf{b}^T&1\end{matrix}\right]$ where $\widetilde{\mathbf{B}}_1 \triangleq \mathbf{bb}^T$, the optimization problem is transformed as 
\begin{subequations}\label{eq:solve for B}\begin{align}
&\underset{\mathbf{U}\in\mathbb{S}_T^{+},\widetilde{\mathbf{B}}}\min
~~\text{Tr}\{\mathbf{U}^{-1}\} + \text{Tr}\{\widetilde{\mathbf{B}}\mathbf{E}_a\}      \\
&\quad\text{s.t.}\quad \left[\begin{matrix} \mathbf{F}_{\bm{\theta}\bm{\theta}^T}(\widetilde{\mathbf{B}}_1)+\mathbf{F}_\theta-\mathbf{U}\bm{\Lambda}& \mathbf{F}_{\bm{\theta}\widetilde{\bm{\alpha}}^T}(\widetilde{\mathbf{B}}_1)\\ \mathbf{F}_{\bm{\theta}\widetilde{\bm{\alpha}}^T}^T(\widetilde{\mathbf{B}}_1)& \mathbf{F}_{\widetilde{\bm{\alpha}}\widetilde{\bm{\alpha}}^T}(\widetilde{\mathbf{B}}_1)+\mathbf{F}_\alpha\end{matrix}\right]\succeq\mathbf{0},\label{eq:cons4b}\\
&\qquad~~T^2\leq\text{Tr}\{\widetilde{\mathbf{B}}_1\mathbf{1}\}\leq (N-K)^2,\\
&\qquad~~0\leq \widetilde{\mathbf{B}}(n,n)\leq 1,~\forall n,\\
&\qquad~~\widetilde{\mathbf{B}}(N+1,N+1) = 1,
\end{align}\end{subequations}
where we define 
\be
\mathbf{E}_a= \left[\begin{matrix}\rho_2\mathbf{I}_N &\rho_2(\mathbf{a}-\mathbf{1}+\bm{\mu}/\rho_2)\\
\rho_2(\mathbf{a}-\mathbf{1}^T+\bm{\mu}/\rho_2)^T&0\end{matrix}\right].
\ee
After solving \eqref{eq:solve for B}, the optimal solution is obtained as 
\be\label{eq:update b}
\mathbf{b} = (\mathbf{1}^T\widetilde{\mathbf{B}}_1\mathbf{1})^{-1/2}\widetilde{\mathbf{B}}_1\mathbf{1},
\ee
or using Gaussian randomization. 

\subsection{Summary}
Based on the above derivations, we summarize the proposed BCRB-oriented joint array partitioning and beamforming design approach in Algorithm~1. With an appropriate initialization, we alternatingly update the beamformer $\mathbf{W}$, the array partitioning vector $\mathbf{a}$, the auxiliary variable $\mathbf{b}$, and the dual variable $\bm{\mu}$ until convergence.

\begin{algorithm}[!t]
    \begin{small}
        \caption{BCRB-Oriented Joint Array Partitioning and Beamforming Design Algorithm}
        \label{alg:1}
        \begin{algorithmic}[1]
            \REQUIRE {$\mathbf{h}_t$, $\forall t$, $\sigma_\text{r}^2$, $\mathbf{h}_k$, $\sigma_k^2$, $\Gamma_k$, $\forall k$, $\mathbf{H}_\text{SI}$, $\mathbf{J}_\tau$, $\bm{\mu}_\theta$, $\bm{\Sigma}_\theta$, $\bm{\mu}_\alpha$, $\bm{\Sigma}_\alpha$, $P$, $L$, $\rho_1$, $\rho_2$.}
            \ENSURE {$\mathbf{a}$, $\mathbf{W}$.}
            \STATE {Initialize $a_n=b_n=0.5,~\forall n$, $\mathbf{R}_\text{n} = \sigma_\text{r}^2\mathbf{I}$, $\bm{\mu} = \mathbf{0}$.}        
            \REPEAT
            \STATE {Obtain $\widetilde{\mathbf{R}}$, $\widetilde{\mathbf{R}}_k,~\forall k$ by solving \eqref{eq:solve for W}.}
            \STATE {Update $\mathbf{W}$ by \eqref{eq:update w} and \eqref{eq:update Wr}.}
            \STATE {Obtain $\widetilde{\mathbf{A}}_1$ by solving \eqref{eq:solve for A}.}
            \STATE{Update $\mathbf{a}$ by \eqref{eq:update a} or Gaussian randomization.}
            \STATE {Obtain $\widetilde{\mathbf{B}}_1$ by solving \eqref{eq:solve for B}.}
            \STATE{Update $\mathbf{b}$ by \eqref{eq:update b} or Gaussian randomization.}
            \STATE{Update $\bm{\mu}$ by \eqref{eq:update mu}.}
            \STATE{Update $\mathbf{R}_\text{n}$ by \eqref{eq:Rn}.}
            \UNTIL {convergence}
            \STATE {Return $\mathbf{a}$, $\mathbf{W}$.}
        \end{algorithmic}
    \end{small}
\end{algorithm}

\section{Joint Array Partitioning and Transmit Beamforming Design for an Extended Target}

In this section, we focus on scenarios involving extended targets that span a contiguous set of angles, such as in vehicular radar applications. Unlike point targets, an extended target reflects signals from multiple scattering points along its surface, resulting in a continuous set of echoes. Accurately modeling and exploiting this structure is crucial for improving parameter estimation and system performance. In the following, 
we first present the extended target (ET) model and formulate the corresponding Bayesian estimation problem. As will be shown, although the system model differs from that of point-like targets, the resulting optimization problem retains a structure similar to that in Section II. Consequently, we adapt the algorithm developed in Section III, introducing appropriate modifications to handle the expressions required for ET modeling. 

\subsection{System Model and Problem Formulation}
The radar returns from an ET collected during $L$ symbol slots at the BS receive array can be expressed as 
\be\label{eq:Yr4ET}
\mathbf{Y}_\text{r} = (\mathbf{I}_N-\mathbf{A})\mathbf{GAWS} +\mathbf{H}_\text{s}\mathbf{WSJ}_\tau + (\mathbf{I}_N-\mathbf{A})\mathbf{N}_\text{r},
\ee
where the target response matrix $\mathbf{G}$ represents the effect of multiple distributed scatters along the ET. In particular, we assume the ET is composed of $N_\text{bins}$ reflectors
from angles $\widehat{\bm{\theta}} \triangleq [\theta_1,\theta_2,\ldots,\theta_{N_\text{bins}}]^T$ with corresponding RCS values $\bm{\alpha} \triangleq [\alpha,\alpha_2,\ldots,\alpha_{N_\text{bins}}]^T$. The target response matrix is given by 
\be
\mathbf{G} = \sum_{i=1}^{N_\text{bins}}\alpha_i\mathbf{h}(\theta_i)\mathbf{h}^T(\theta_i),
\ee
where $\mathbf{h}(\theta_i)$ represents the LoS channel for the $i$-th scatterer at DOA $\theta_i$. 

Unlike multiple point-like targets, an ET can be efficiently parameterized by its central angle $\theta_\text{c}$ and angular spread $\Delta_\theta$, as introduced in \cite{RLiu-arXiv-2025}. Specifically, the scatterer angles are modeled as 
\begin{equation}
\label{wshift}
    \widehat{\bm{\theta}} = \theta_\text{c} + \Delta_\theta\mathbf{w},
\end{equation}
where $\mathbf{w}\triangleq [w_1,w_2,\ldots,w_{N_\text{bins}}]^T \subset [-1,1]$ captures the spatial distribution of the scatterers and is determined by the physical properties of the ET. Consequently, the parameters to be estimated for this scenario are
\begin{equation}
    \bm{\xi} \triangleq [\bm{\theta},~ \Re\{\bm{\alpha}^T\},~ \Im\{\bm{\alpha}^T\}]^T,
\end{equation}
where $\bm{\theta} \triangleq [\theta_\text{c},~ \Delta_\theta]^T$ contains the central angle and angular spread that describe the ET in the spatial domain, and $\bm{\alpha}$ represents the RCS parameters for each angle bin.

Given the received signals: $\widetilde{\mathbf{y}}_\text{r} = \text{vec}\{\mathbf{Y}_\text{r}\} \sim \mathcal{CN}(\bm{\eta}, \widetilde{\mathbf{R}}_\text{n})$ with $\bm{\eta} = \text{vec}\{(\mathbf{I}-\mathbf{A})\mathbf{GAWS}\}$, the likelihood function for estimating $\bm{\xi}$ has the same general form as in \eqref{eq:LLF}, but with $\bm{\eta}$ defined by the ET model. As before, the likelihood-based FIM is given by $\mathbf{F}_\text{L}(i,j) = 2\Re\left\{\frac{\partial \bm{\eta}^H}{\partial \xi_i} \widetilde{\mathbf{R}}_\text{n}^{-1} \frac{\partial \bm{\eta}}{\partial \xi_j}\right\}$, in this case with partial derivatives  
\begin{subequations}\begin{align}
 \frac{\partial \bm{\eta}}{\partial \theta_\text{c}} &= \text{vec}\left\{\mathbf{BH}_\theta\mathbf{AWS}\right\}, \\
 \frac{\partial \bm{\eta}}{\partial \Delta_\theta} &= \text{vec}\left\{\mathbf{BH}_\Delta\mathbf{AWS}\right\},\\
 \frac{\partial \bm{\eta}}{\partial \Re\{\alpha_i\}} &= \text{vec}\{\mathbf{BH}_i \mathbf{AWS}\},\\
 \frac{\partial \bm{\eta}}{\partial \Im\{\alpha_i\}} &= \jmath \text{vec}\{\mathbf{BH}_i \mathbf{AWS}\},
\end{align}\end{subequations}
where 
\begin{subequations}\begin{align}
\mathbf{B}&\triangleq\mathbf{I}-\mathbf{A},\qquad \mathbf{H}_i \triangleq \mathbf{h}(\theta_i)\mathbf{h}^T(\theta_i),\\
 \mathbf{H}_\theta &\triangleq \frac{\partial \mathbf{G}}{\partial \theta_\text{c}} = -\jmath\pi\sum_{i=1}^{N_\text{bins}}\alpha_i \cos\theta_i(\mathbf{QH}_i+\mathbf{H}_i\mathbf{Q}),\\
\mathbf{H}_\Delta &\triangleq \frac{\partial \mathbf{G}}{\partial \Delta_\theta} = -\jmath\pi\sum_{i=1}^{N_\text{bins}}\alpha_i w_i\cos\theta_i(\mathbf{QH}_i+\mathbf{H}_i\mathbf{Q}).
\end{align}\end{subequations}
The elements of $\mathbf{F}_\text{L}$ for the ET model can then be computed as shown in \eqref{eq:FL ele} on the top of the next page.

In addition, we assume the following prior distributions for the central angle $\theta_\text{c}$, the angular spread $\Delta_\theta$, and the RCS $\bm{\alpha}$:
\be
\theta_\text{c}\sim\mathcal{N}(\mu_\text{c},\sigma_\text{c}^2),~~ \Delta_\theta\sim\mathcal{N}(\mu_\Delta,\sigma^2_\Delta),~~
\bm{\alpha} \sim \mathcal{CN}(\bm{\mu}_\alpha, \bm{\Sigma}_\alpha).
\ee
For simplicity, we use the notation $\bm{\theta} \triangleq[\theta_\text{c},\Delta_\theta]^T$ and rewrite the prior distribution of $\bm{\theta}$ as
\be
\bm{\theta}\sim\mathcal{N}(\bm{\mu}_\theta,\bm{\Sigma}_\theta),
\ee
where $\bm{\mu}_\theta = [\mu_\text{c},\mu_\Delta]^T$ and  $\bm{\Sigma}_\theta = \left[\begin{matrix}
    \sigma_\text{c}^2 & 0\\ 0 & \sigma_\Delta^2
\end{matrix}\right]$. The corresponding prior probabilities and the prior FIM follow the same form as in \eqref{eq:prior prob} and \eqref{eq:Fp partition}.

Following the Bayesian framework, the FIM is computed as $\mathbf{F}_\text{B}=\mathbf{F}_\text{L}+\mathbf{F}_\text{P}$ and the BCRB is obtained by inverting $\mathbf{F}_\text{B}$. To obtain a closed-form expression for the BCRB of the central angle $\theta_\text{c}$ and the angular spread $\Delta_\theta$, we again partition the FIM matrices $\mathbf{F}_\text{L}$ and $\mathbf{F}_\text{P}$ into $2\times 2$ blocks, similar to \eqref{eq:FL partition} and \eqref{eq:Fp partition}. In the given ET scenario, the dimension of $\mathbf{F}_{\bm{\theta\theta}^T}$ and $\mathbf{F}_\theta$ is $2\times 2$, $\mathbf{F}_{\bm{\theta}\widetilde{\bm{\alpha}}^T}$ is $2\times 2N_\text{bins}$, $\mathbf{F}_{\widetilde{\bm{\alpha}}\widetilde{\bm{\alpha}}^T}$ and $\mathbf{F}_\alpha$ is $2N_\text{bins}\times 2N_\text{bins}$, respectively. 
Consequently, the weighted BCRB for estimating $\theta_\text{c}$ and $\Delta_\theta$ follows the same form as \eqref{eq:objective}, leading to an optimization problem analogous to \eqref{eq:original problem}.

\begin{figure*}
\begin{subequations}\label{eq:FL ele}
\begin{align}\label{eq:FL11}
    \mathbf{F}_\text{L}(1,1)&= 2L\Re\{\text{Tr}\{\mathbf{B}\mathbf{R}_\text{n}^{-1}\mathbf{B}\mathbf{H}_\theta\mathbf{AR}_\text{w}\mathbf{A}\mathbf{H}^H_\theta\}\} ,\\
    \mathbf{F}_\text{L}(1,2)&= 2L\Re\{\text{Tr}\{\mathbf{B}\mathbf{R}_\text{n}^{-1}\mathbf{B}\mathbf{H}_\Delta\mathbf{AR}_\text{w}\mathbf{A}\mathbf{H}^H_\theta\}\} ,\\
    \mathbf{F}_\text{L}(2,2)&= 2L\Re\{\text{Tr}\{\mathbf{B}\mathbf{R}_\text{n}^{-1}\mathbf{B}\mathbf{H}_\Delta\mathbf{AR}_\text{w}\mathbf{A}\mathbf{H}^H_\Delta\}\} ,\\
    \mathbf{F}_\text{L}(1,i+2)&= 2L\Re\{\text{Tr}\{\mathbf{B}\mathbf{R}_\text{n}^{-1}\mathbf{B}\mathbf{H}_i\mathbf{AR}_\text{w}\mathbf{A}\mathbf{H}^H_\theta\}\}  ,\\ 
    \mathbf{F}_\text{L}(1,N_\text{bins}+i+2)&= 2L\Re\{\jmath\text{Tr}\{\mathbf{B}\mathbf{R}_\text{n}^{-1}\mathbf{B}\mathbf{H}_i\mathbf{AR}_\text{w}\mathbf{A}\mathbf{H}^H_\theta\}\} ,\\
        \mathbf{F}_\text{L}(2,i+2)&= 2L\Re\{\text{Tr}\{\mathbf{B}\mathbf{R}_\text{n}^{-1}\mathbf{B}\mathbf{H}_i\mathbf{AR}_\text{w}\mathbf{A}\mathbf{H}^H_\Delta\}\} ,\\ 
        \mathbf{F}_\text{L}(2,N_\text{bins}+i+2)&= 2L\Re\{\jmath\text{Tr}\{\mathbf{B}\mathbf{R}_\text{n}^{-1}\mathbf{B}\mathbf{H}_i\mathbf{AR}_\text{w}\mathbf{A}\mathbf{H}^H_\Delta\}\} ,\\ 
    \mathbf{F}_\text{L}(i+2,j+2)&= 2L\Re\{\text{Tr}\{\mathbf{B}\mathbf{R}_\text{n}^{-1}\mathbf{B}\mathbf{H}_j\mathbf{AR}_\text{w}\mathbf{A}\mathbf{H}_i^H\}\} ,\\    
        \mathbf{F}_\text{L}(N_\text{bins}+i+2,j+2)&= 2L\Re\{-\jmath\text{Tr}\{\mathbf{B}\mathbf{R}_\text{n}^{-1}\mathbf{B}\mathbf{H}_j\mathbf{AR}_\text{w}\mathbf{A}\mathbf{H}_i^H\}\} ,\\ 
      \mathbf{F}_\text{L}(i+2,N_\text{bins}+j+2)&= 2L\Re\{\jmath\text{Tr}\{\mathbf{B}\mathbf{R}_\text{n}^{-1}\mathbf{B}\mathbf{H}_j\mathbf{AR}_\text{w}\mathbf{A}\mathbf{H}_i^H\}\},\\ 
       \mathbf{F}_\text{L}(N_\text{bins}+i+2,N_\text{bins}+j+2)&= 2L\Re\{\text{Tr}\{\mathbf{B}\mathbf{R}_\text{n}^{-1}\mathbf{B}\mathbf{H}_j\mathbf{AR}_\text{w}\mathbf{A}\mathbf{H}_i^H\}\}. 
\end{align}\end{subequations}  
\rule[-0pt]{18.5 cm}{0.05em}
\end{figure*}

\subsection{Joint Array Partitioning and Beamforming Design}

Despite the differences in the ET model, the resulting optimization problem maintains a structure analogous to the point-like target case, differing  primarily in the expressions derived from \eqref{eq:FL ele}. Therefore, the algorithm proposed in Section III can be applied here with only slight modifications to accommodate the new FIM expressions. Following the procedures in Sections III-A, III-B, and III-C, we convert the weighted BCRB minimization into a multivariate optimization problem similar to \eqref{eq:ALp}, noting that $\mathbf{U}\in\mathbb{S}_2^+$ and $\mathbf{F}_\text{L}$ is replaced by \eqref{eq:FL ele}. Then, the dual variable $\bm{\mu}$ is updated via \eqref{eq:update mu}, and the variables $\mathbf{W}$, $\mathbf{a}$, and $\mathbf{b}$ are updated using the SDR approach presented in Sections III-D, III-E, and III-F, respectively, except that the FIM expressions in \eqref{eq:constraint in Rw}, \eqref{eq:cons4a}, and \eqref{eq:cons4b} must be replaced by their extended-target counterparts in \eqref{eq:FL ele}.

To illustrate how these expressions can be cast into the same form required by our algorithm, we take the first element $\mathbf{F}_\text{L}(1,1)$ in \eqref{eq:FL11} as an example. Define $\mathbf{E}_{1,1}\triangleq \mathbf{H}_\theta^H\mathbf{B}\mathbf{R}_\text{n}^{-1}\mathbf{B}\mathbf{H}_\theta$ and $\mathbf{D}_{1,1}\triangleq \mathbf{H}_\theta\mathbf{AR}_\text{w}\mathbf{A}\mathbf{H}^H_\theta$. We can rewrite 
\begin{subequations}\begin{align}
\mathbf{F}_\text{L}(1,1)&=\text{Tr}\{\mathbf{AR}_\text{w}\mathbf{A}\mathbf{E}_{1,1}\} = \sum_{i=1}^{N+K}\!\text{Tr}\{\mathbf{Aw}_i\mathbf{w}_i^H\mathbf{AE}_{1,1}\} \non \\
&= \sum_{i=1}^{N+K}\text{Tr}\big\{\widetilde{\mathbf{A}}_1\text{diag}\{\mathbf{w}_i^H\}\mathbf{E}_{1,1}\text{diag}\{\mathbf{w}_i\}\big\},\\
\mathbf{F}_\text{L}(1,1)&=\text{Tr}\{\mathbf{B}\mathbf{R}_\text{n}^{-1}\mathbf{BD}_{1,1}\} = \sum_{n=1}^{N}\text{Tr}\{\mathbf{Bu}_n\mathbf{u}_n^H\mathbf{BD}_{1,1}\} \non \\
&= \sum_{n=1}^N\text{Tr}\big\{\widetilde{\mathbf{B}}_1\text{diag}\{\mathbf{u}_n^H\}\mathbf{D}_{1,1}\text{diag}\{\mathbf{u}_n\}\big\},
\end{align}\end{subequations}
where we use an eigendecomposition to construct $\mathbf{R}_\text{n}^{-1} = \sum_{n=1}^N\mathbf{u}_n\mathbf{u}_n^H$. Using the same approach for other elements of the ET-based FIM, the algorithm introduced in Section III can be seamlessly employed in this setting.

\section{Joint Maximum A Posteriori Estimation}

Building on the optimization framework developed for both multiple point-like target and ET scenarios, this section presents an approach for joint estimation of the DOAs and RCSs. We formulate the estimation problem within a Bayesian framework, wherein the target parameters are inferred by maximizing the posterior distribution. This formulation effectively merges prior statistical information with the observed data, thereby improving estimation accuracy and robustness.

To facilitate the estimation process in the multiple point-like targets scenario, we first rewrite the received signals in \eqref{eq: yr1} in a compact form as 
\begin{align}
 \widetilde{\mathbf{y}}_\text{r} = \mathbf{V}(\bm{\theta})\bm{\alpha} + \mathbf{n},
\end{align}
where we define 
\begin{equation}
\mathbf{V}(\bm{\theta}) \triangleq 
\big[\text{vec}\{\mathbf{H}_1 \mathbf{WS}\},\text{vec}\{\mathbf{H}_2 \mathbf{WS}\},\cdots,\text{vec}\{\mathbf{H}_T \mathbf{WS}\}\big].  
\end{equation}
In the Bayesian framework, the negative log-posterior for parameter estimation given in \eqref{eq:J} can be expressed as 
\begin{equation}\begin{aligned}
\mathcal{L}(\bm{\theta}, \bm{\alpha}) &= (\widetilde{\mathbf{y}}_\text{r} - \mathbf{V}(\bm{\theta}) \bm{\alpha})^H \widetilde{\mathbf{R}}_\text{n}^{-1} (\widetilde{\mathbf{y}}_\text{r} - \mathbf{V}(\bm{\theta}) \bm{\alpha}) \\
&\quad~ + \frac{1}{2} (\bm{\theta} - \bm{\mu}_\theta)^T \bm{\Sigma}_\theta^{-1} (\bm{\theta} - \bm{\mu}_\theta)\\
&\quad~ + (\bm{\alpha} - \bm{\mu}_\alpha)^H \bm{\Sigma}_\alpha^{-1} (\bm{\alpha} - \bm{\mu}_\alpha),
\end{aligned}\end{equation}
which is minimized to obtain the desired parameter estimates. 
Since the model depends linearly on the RCS parameters $\bm{\alpha}$, the estimation can be simplified by first solving for $\bm{\alpha}$ with $\bm{\theta}$ fixed. Setting the gradient of $\mathcal{L}(\bm{\theta},\bm{\alpha})$ to zero, 
the optimal closed-form solution for $\bm{\alpha}$ is obtained as 
\begin{equation}\label{eq:alpha est}
\hat{\bm{\alpha}} = \mathbf{D}^{-1} \mathbf{p},
\end{equation}
where 
\begin{equation}\label{eq:Dp}
\begin{aligned}
 \mathbf{D} &= \mathbf{V}^H(\bm{\theta}) \widetilde{\mathbf{R}}_\text{n}^{-1} \mathbf{V}(\bm{\theta}) + \bm{\Sigma}_\alpha^{-1}, \\
\mathbf{p} &= \mathbf{V}^H(\bm{\theta}) \widetilde{\mathbf{R}}_\text{n}^{-1} \widetilde{\mathbf{y}}_\text{r} + \bm{\Sigma}_\alpha^{-1} \bm{\mu}_\alpha.   
\end{aligned}\end{equation}

Substituting the closed-form expression for $\hat{\bm{\alpha}}$ into the cost function yields a concentrated criterion that depends only on $\bm{\theta}$:
\begin{equation}\label{eq:J2theta}
\mathcal{L}(\bm{\theta}) = -\mathbf{p}^H \mathbf{D}^{-1} \mathbf{p} + \frac{1}{2} (\bm{\theta} - \bm{\mu}_\theta)^T \bm{\Sigma}_\theta^{-1} (\bm{\theta} - \bm{\mu}_\theta).
\end{equation}
Due to the nonlinearity of $\mathcal{L}(\bm{\theta})$ with respect to $\bm{\theta}$, we employ Newton's method to minimize \eqref{eq:J2theta}, leading to the following update at the $m$-th iteration:
\begin{equation}\label{eq:update theta}
\bm{\theta}^{(m+1)} = \bm{\theta}^{(m)}-d^{(m)}\big(\nabla^2 \mathcal{L}(\bm{\theta}^{(m)})\big)^{-1}\nabla \mathcal{L}(\bm{\theta}^{(m)}),
\end{equation}
where $\bm{\theta}^{(m)}$ is the current estimate and $d^{(m)}$ is the step size computed according to the Armijo rule. The gradient $\nabla \mathcal{L}$ and the Hessian matrix $\nabla^2 \mathcal{L}$ are derived in Appendix A. The complete estimation procedure is summarized in Algorithm~2.

For the ET scenario, the same algorithm can be adapted to estimate the central angle $\theta_\text{c}$, the angular spread $\Delta_\theta$, and the RCS $\bm{\alpha}$. However, in this case, $\mathbf{V}(\bm{\theta})$ depends only on $\theta_\text{c}$ and $\Delta_\theta$, with a different formulation of the gradient and Hessian matrices, as presented in Appendix A. Consequently, the resulting approach provides a unified framework that accommodates both point-like and extended targets by appropriately adjusting the signal model and its associated derivatives.

\begin{algorithm}[!t]
    \begin{small}
        \caption{Newton Method-Based Joint MAP Estimation Algorithm}
        \label{alg:1}
        \begin{algorithmic}[1]
            \REQUIRE {$\widetilde{\mathbf{y}}_\text{r}$,  $\mathbf{a}$, $\mathbf{WS}$.}
            \ENSURE {$\bm{\theta}$, $\bm{\alpha}$.}
            \STATE {Initialize $\bm{\theta} = \bm{\mu}_\theta$, $\bm{\alpha} = \bm{\mu}_\alpha$, and $m=0$.}        
            \REPEAT
            \STATE {Calculate $\mathbf{D}$ and $\mathbf{p}$ by \eqref{eq:Dp}.}
            \STATE {Calculate the gradient $\nabla \mathcal{L}$ by \eqref{eq:gradient}.}
            \STATE {Calculate the Hessian $\nabla^2 \mathcal{L}$ by \eqref{eq:Hessian}.}
            \STATE{Set $\nabla^2 \mathcal{L}=\mathbf{I}$ when $\nabla^2 \mathcal{L}$ is not semidefinite. }
            \STATE {Determine the step size $d^{(m)}$ using the Armijo rule.}
            \STATE {Update $\bm{\theta}$ by \eqref{eq:update theta}.}    
            \STATE {Set $m:= m+1$.}
            \UNTIL {Convergence criterion is satisfied.}
            \STATE {Obtain $\bm{\alpha}$ using \eqref{eq:alpha est}. }
            \STATE {Return $\bm{\theta}$ and $\bm{\alpha}$.}
        \end{algorithmic}
    \end{small}
\end{algorithm}

\section{Simulation Results}
In this section, we evaluate the benefits of the proposed joint array partitioning and beamforming designs through extensive simulation results. For the simulations, we adopt a Rician fading channel model for the communication users, which is mathematically expressed as 
\begin{equation}
    \mathbf{h}_k = \sqrt{\frac{\kappa}{1+\kappa}}\mathbf{h}_k^{\text{LoS}} + \sqrt{\frac{1}{1+\kappa}}\mathbf{h}_k^{\text{NLoS}},
\end{equation}
where the LoS component is defined as $\mathbf{h}_k^{\text{LoS}}\triangleq[e^{-\jmath\frac{1-N}{2}\pi\sin\phi_k}, e^{-\jmath\frac{3-N}{2}\pi\sin\phi_k},\ldots,e^{-\jmath\frac{N-1}{2}\pi\sin\phi_k}]^T$, $\phi_k$ represents the azimuth angle of the $k$-th user, and the elements of the NLoS component $\mathbf{h}_k^{\text{NLoS}}\in\mathbb{C}^{N}$ are drawn from $\mathcal{CN}(0,1)$. The Rician factor is set to $\kappa=3$dB, and the standard distance-dependent path loss model $\text{PL}(d) = C_0(d/d_0)^{-\beta}$ is adopted, where $d$ denotes the link distance, $\beta$ is the path-loss exponent, and $C_0=-30$ dB represents the reference path loss at a distance of $d_0=1$ m. The distances for the BS-target and BS-user links are set to $50$m and $100$m, respectively, with corresponding path-loss exponents of 2 and 2.6. The self-interference channel is modeled as $\mathbf{H}_\text{SI}(i,j) = \alpha_\text{SI}e^{-\jmath 2\pi d_{i,j}/\lambda}$, where $\alpha_\text{SI}$ represents the amplitude of the residual self-interference after cancellation, $d_{i,j}$ is the distance between the $i$-th and the $j$-th antennas, and $\lambda$ is the signal wavelength. We set the carrier frequency to $3.5$GHz with a bandwidth of $100$MHz. Leveraging advanced analog- and digital-domain cancellation techniques developed for full-duplex systems \cite{ASabharwal-JSAC-2014}, \cite{MErdem-WC-2021}, we assume that the strength of the self-interference channel is comparable to that of the target echoes. Without loss of generality, the noise power at all receivers is set to $\sigma_k^2=\sigma_\text{r}^2 = -80$ dBm. The SINR requirements for all users are set to a uniform value of $\Gamma_k = 10$ dB for all $k$, and the weighting matrix in \eqref{eq:objective} is assumed to be $\bm{\Lambda} = \mathbf{I}$. The number of collected samples within one coherent processing interval is set to $L = 32$.

\begin{figure}[!t] \centering\vspace{-0.3 cm}
\includegraphics[width = 0.9\linewidth]{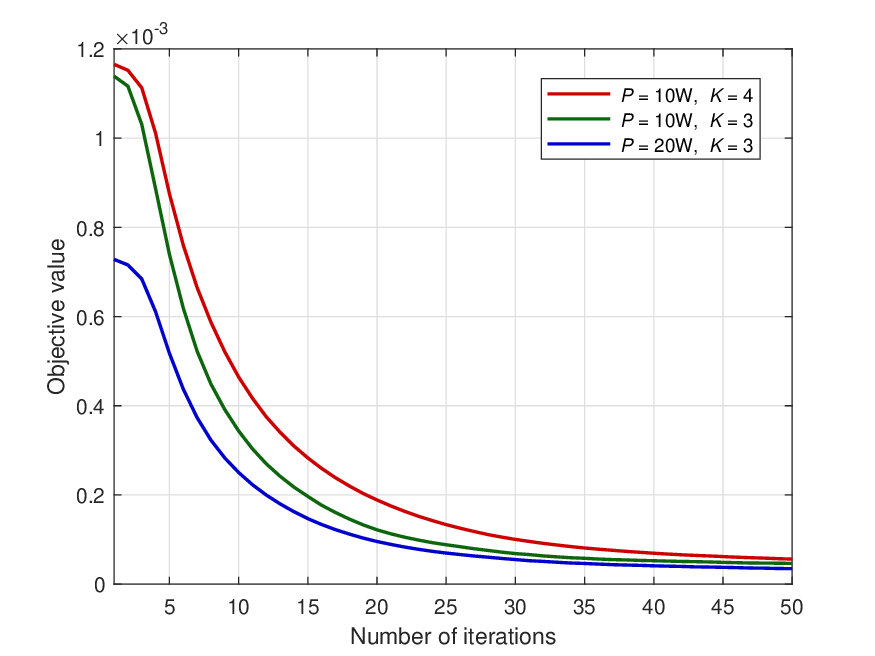}	
\caption{Convergence performance of Algorithm 1.}\label{fig:conv1}\vspace{-0.3 cm}
\end{figure}

To demonstrate the effectiveness of the proposed joint array partitioning scheme (denoted as ``\textbf{Prop.}''), we compare its performance against two benchmark approaches that assume fixed array partitions. The first, referred to as ``\textbf{Even}'', is commonly employed in the ISAC literature and partitions the array into equally sized contiguous transmit and receive subarrays: $\mathbf{a} = [\underbrace{1~1~\ldots~1}_{N/2}~\underbrace{0~0~\ldots~0}_{N/2}]^T$. While this scheme offers a straightforward and intuitive partitioning strategy, it fails to fully exploit the array's potential for balancing joint sensing and communication. The second benchmark follows a heuristic approach aimed at maximizing the array aperture for enhanced sensing performance. In this approach, referred to as ``\textbf{Heu.}'', the receive antennas are located at both ends of the array: $\mathbf{a} = [\underbrace{1~1\ldots1}_{N/4}~\underbrace{0~0\ldots0}_{N/2}~\underbrace{1~1\ldots~1}_{N/4}]^T$. This heuristic approach is effective for improving sensing capabilities, particularly in single-target detection scenarios \cite{RLiu-Asilomar-2024}, but we will see that it does not achieve optimal performance when balancing both sensing and communication functionalities, especially in complex environments. 
The proposed method dynamically optimizes the array partitioning to achieve a balance between communication and sensing objectives, yielding performance that surpasses these fixed partitioning strategies.
 
\begin{figure}[!t] \centering	
\includegraphics[width = 0.9\linewidth]{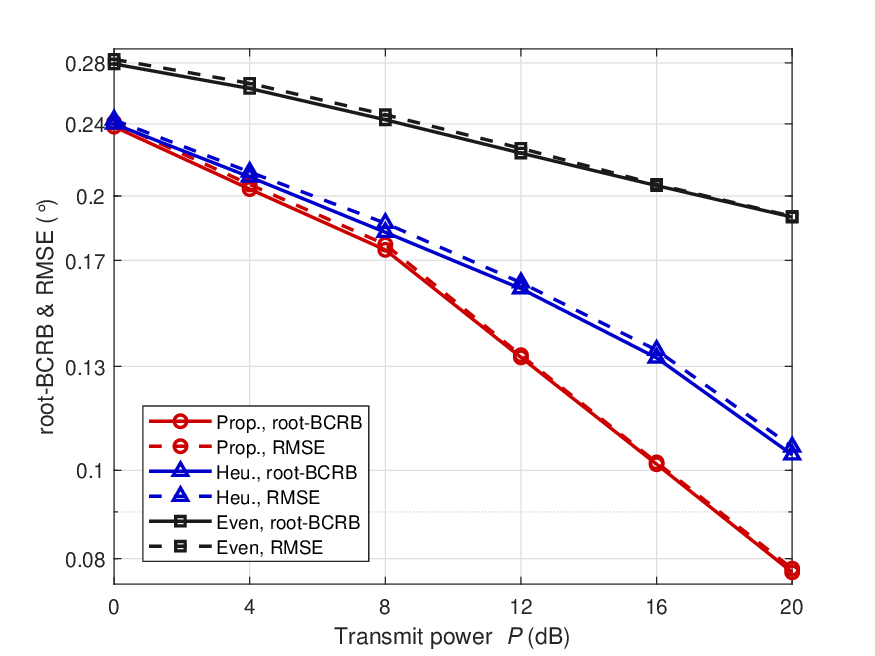}	
\caption{Root-BCRB and RMSE versus transmit power $P$.}\label{fig:power}\vspace{-0.3 cm}
\end{figure}

\subsection{Multiple Point-Like Targets Scenarios}

For the case of multiple point targets, we consider a dual-functional BS equipped with a ULA consisting of $N=24$ elements, serving $K=3$ communication users and sensing $T=3$ targets. The mean DOAs and RCSs of the targets are set as $\bm{\mu}_\theta = [50^\circ, 60^\circ, 70^\circ]^T$ and $\bm{\mu}_\alpha = \sqrt{2}e^{\jmath\pi/4}$, respectively. The variance of the priors is $\bm{\Sigma}_\theta = 0.09\mathbf{I}_3$ for the DOAs and $\bm{\Sigma}_\alpha = 0.01\mathbf{I}_3$ for the RCSs. Fig.~\ref{fig:conv1} illustrates the convergence behavior of the proposed algorithm by plotting the objective of \eqref{eq:obj1} against the number of iterations under various settings. We see that convergence is achieved in all scenarios within 30 iterations, highlighting the efficiency of the proposed algorithm. 

\begin{figure}[t] 
\centering
\includegraphics[width = 0.9\linewidth]{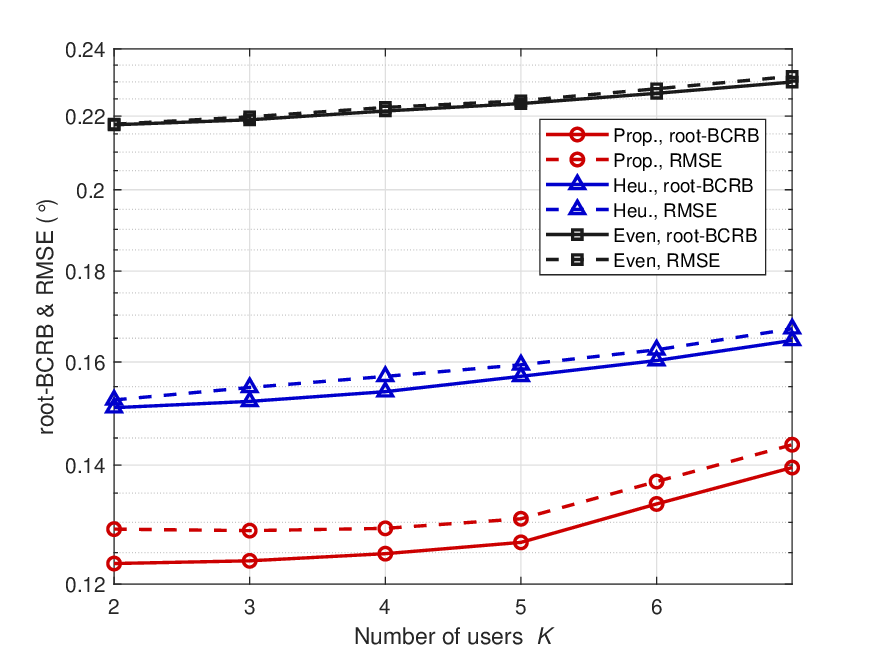}	
\caption{Root-BCRB and RMSE versus number of users $K$.}\label{fig:K}\vspace{-0.3 cm}
\end{figure}

\begin{figure}[t]
\centering
\includegraphics[width = 0.9\linewidth]{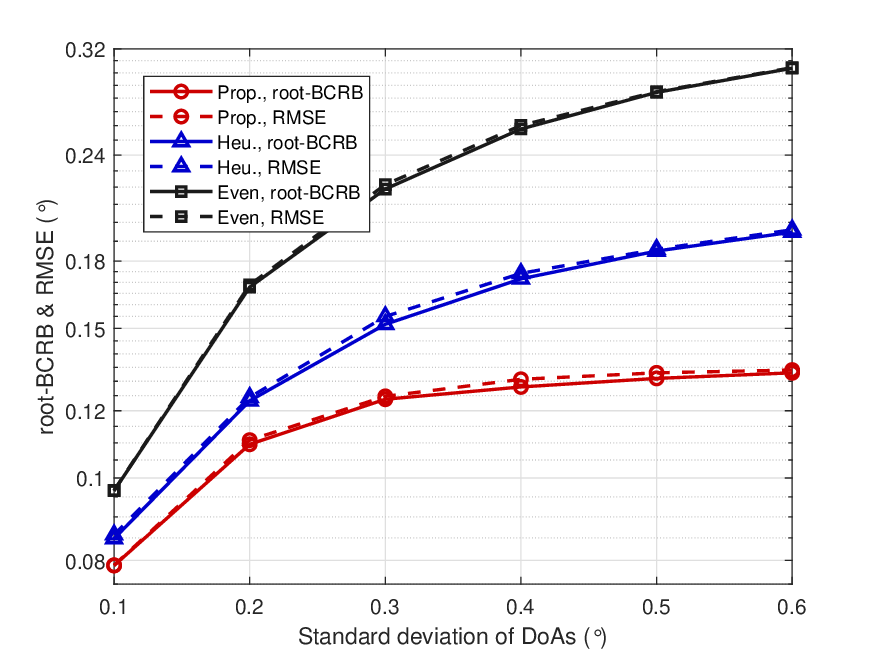}	
\caption{Root-BCRB and RMSE versus standard deviation of DOAs.}\label{fig:var}\vspace{-0.3 cm}
\end{figure}

\begin{figure}[t]	
\centering
\includegraphics[width = 0.9\linewidth]{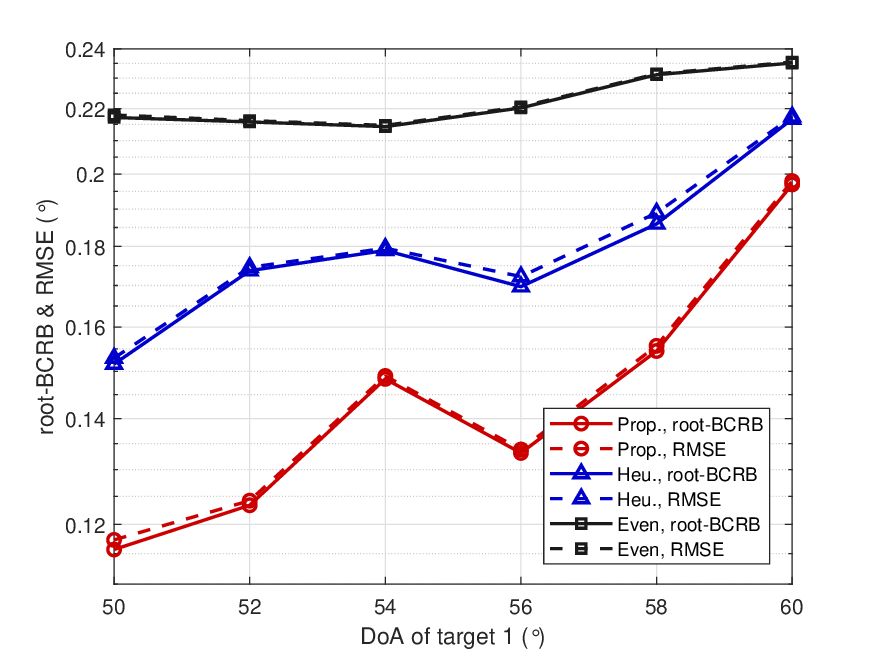}	
\caption{Root-BCRB and RMSE versus the DOA of target 1.}\label{fig:DoA}\vspace{-0.3 cm}
\end{figure}

\begin{figure}[!t]	
\centering
\includegraphics[width = 0.9\linewidth]{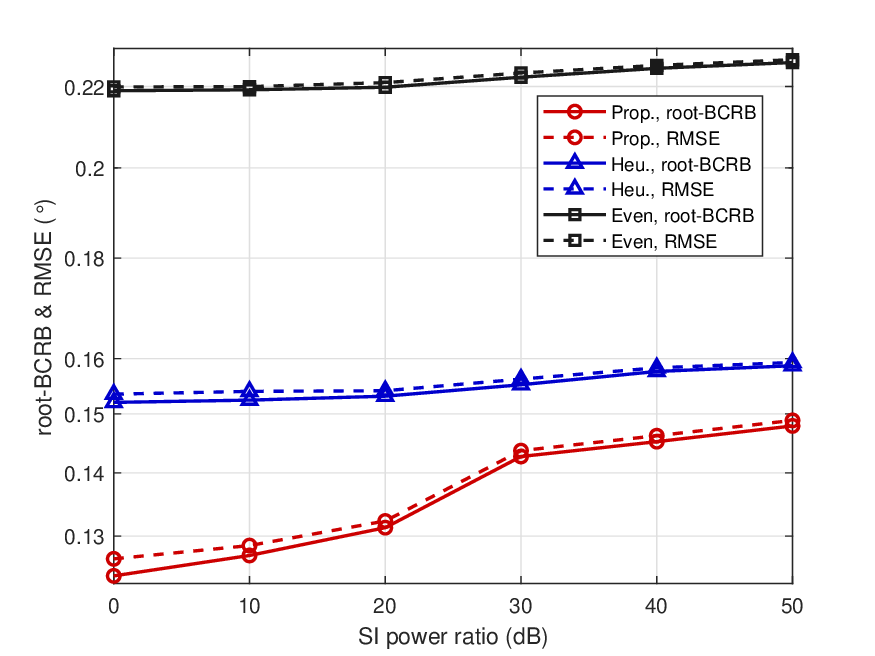}	
\caption{Root-BCRB and RMSE versus the SI channel strength.}\label{fig:SI}
\vspace{-0.3 cm}
\end{figure}

In Figs.~\ref{fig:power}-\ref{fig:SI}, we evaluate the sensing performance in terms of the average of root-BCRB and RMSE across the $T$ targets.
The sensing performance versus transmit power is plotted in Fig.~\ref{fig:power}, where solid lines denote the root-BCRB, representing the theoretical lower bounds determined by the array partitioning and transmit beamforming, and the dashed lines correspond to the root MSE of the estimates obtained by the proposed MAP estimation algorithm in Section V. The root MSE of all methods aligns well with the root-BCRB for different power levels, validating the effectiveness of the MAP estimation algorithm. More importantly, the proposed array partitioning scheme significantly outperforms the two benchmarks, especially in the high-power region. The root MSE of the proposed algorithm is approximately 75\% less than that of the ``Heu.'' scheme and 50\% of the ``Even'' partitioning at $P=20$ dB. This substantial improvement underscores the benefit of dynamic array partitioning architecture.

Fig.~\ref{fig:K} presents the sensing performance as a function of the number of users, highlighting the inherent trade-off between target DOA estimation accuracy and multiuser communication demands. The proposed algorithm based on dynamic array partitioning consistently outperforms the benchmarks for all scenarios, demonstrating its effectiveness in balancing 
the competing ISAC objectives. As the communication requirements become more stringent, the dynamic partitioning continues to provide significant improvements in sensing performance, although the additional gains gradually diminish due to increasing resource allocation constraints.

Fig. \ref{fig:var} examines the impact of the prior standard deviation of the target DOAs on sensing performance. As the prior information becomes less reliable, both the root-BCRB and root MSE increase due to increased initial uncertainty about the DOAs, but the improvement provided by the observations increases significantly, as evidenced by the increasing difference between the RMSE and the initial prior DOA standard deviation. The gain offered by the proposed dynamic array partitioning is more pronounced as the initial uncertainty increases, further indicating the benefits of the proposed approach in ensuring robust sensing performance.

The resolution performance of the algorithm is illustrated in Fig.~\ref{fig:DoA}, where target 1 is moved from $50^\circ$ to $60^\circ$, while the other two targets remain fixed at $60^\circ$ and $70^\circ$. As target 1 approaches the fixed targets, both the root-BCRB and root MSE increase due to heightened interference resulting from reduced angular separation. However, an interesting dip in both metrics appears around $56^\circ$, where a more favorable array geometry can be achieved that momentarily increases spatial resolution and helps mitigate interference. The proposed algorithm achieves a substantially larger reduction in estimation error compared to the benchmarks, underscoring its ability to adapt the array configuration for improved spatial resolution.

The impact of self-interference on sensing performance is evaluated in Fig.~\ref{fig:SI}, where the SI channel power ratio is defined as $\|\mathbf{H}_\text{SI}\|_F^2/\|\mathbf{h}_t\mathbf{h}_t^T\|_F^2$. The results indicate that the proposed scheme achieves greater performance gains for SIs at 20dB or lower, but the gains diminish as the SI becomes increasingly strong. This results because the proposed dynamic antenna allocation allows the transmit and receive antennas to be positioned in closer proximity than the benchmarks, which maintain non-overlapping and spatially separated transmit and receive arrays. Consequently, the reduced antenna separation in the proposed approach makes it more susceptible to self-interference. These results highlight the necessity of achieving sufficient self-interference cancellation in dynamic array partitioning scenarios.

\begin{figure}[t]	
\centering
\includegraphics[width = 0.9\linewidth]{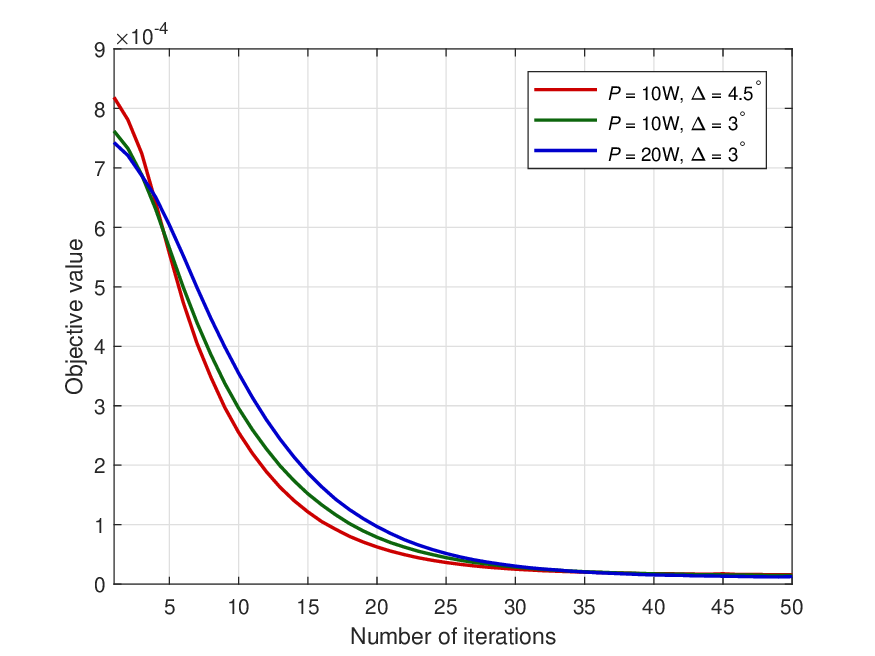}	
\caption{Convergence performance of Algorithm 2.}\label{fig:conv2}\vspace{-0.3 cm}
\end{figure}

 \begin{figure}[t]
 \centering
\includegraphics[width = 0.9\linewidth]{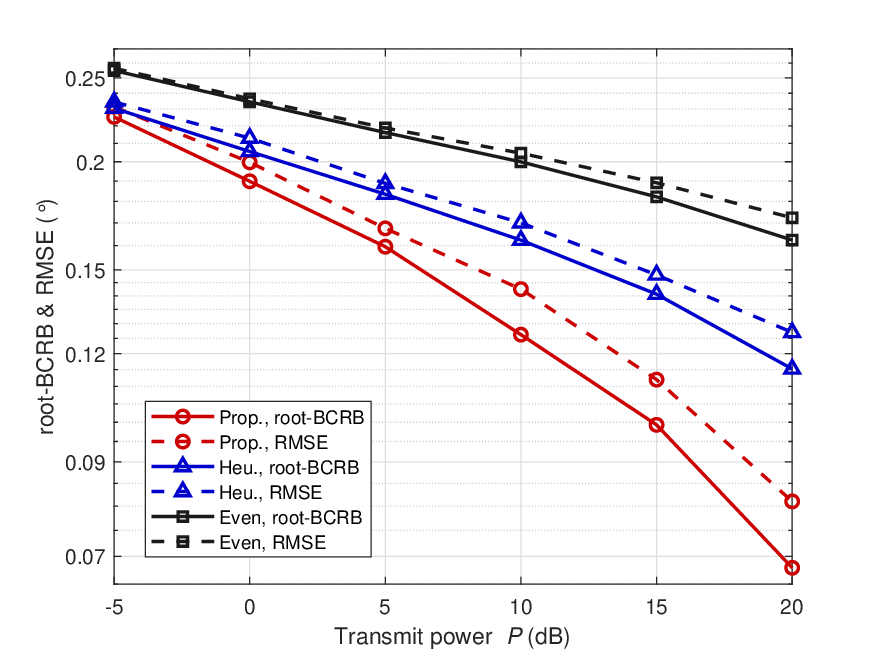}	
\caption{Root-BCRB and RMSE versus transmit power $P$.}\label{fig:P_ET}\vspace{-0.3 cm}
\end{figure}

\vspace{-0.3 cm}
\subsection{Extended Target Scenarios}

In this subsection, we evaluate the performance of the approach proposed in Section IV for ET scenarios. We assume a BS equipped with a ULA of $N = 32$ elements serving $K=3$ communication users and sensing an ET that is modeled with $N_\text{bins} = 5$ reflection points. The positions of these points determined by the vector $\mathbf{w}$ in \eqref{wshift} are assumed to be uniformly sampled in the range $[-1,1]$. 
We consider a typical automotive radar scenario where the ET is located $50$ meters from the BS, beyond the near-field region of the BS array. The ET is located at an azimuth angle of $30^\circ$ and is assumed to have a horizontal dimension of $5.24$m, corresponding to an angular spread of $3^\circ$. For an equivalent MIMO radar system with 16 transmit and 16 receive antennas, the achievable angular resolution is $0.45^\circ$, making it feasible to accurately extract information from $N_\text{bins}=5$ reflection points.
The mean values of the prior distributions for the central angle, angular spread, and RCSs are set to $\mu_\text{c} = 30^\circ$, $\mu_\Delta = 3^\circ$, and $\bm{\mu}_\alpha = \sqrt{2}e^{\jmath\pi/4}$, respectively. The corresponding variances are $\sigma_\text{c}^2=\sigma_\Delta^2=0.09$ and $\bm{\Sigma}_\alpha = \mathbf{I}$, respectively. 

The convergence performance of the proposed algorithm for the ET scenario is shown in Fig.~\ref{fig:conv2}, which shows similar performance as in Fig.~\eqref{fig:conv1} for the point-like target case.  Fig.~\ref{fig:P_ET} presents the sensing performance as a function of the transmit power, where the solid lines represent the theoretical bounds and the dashed lines correspond to the actual estimation error obtained using the joint MAP estimation algorithm in Section V. The root BCRB and RMSE plots show the average of the bound for the central angle and angular spread. The proposed array partitioning scheme clearly provides significant performance gains thanks to its spatial flexibility. In addition, a more obvious gap is seen between the BCRB and the actual estimation error when compared to the multiple point-like target scenario in Fig. \ref{fig:power}. This deviation arises primarily from the stronger nonlinearities introduced by the ET model, which challenge the second-order approximation used in the Newton-based MAP algorithm. These results underscore the necessity of developing more sophisticated estimation algorithms to better address the complexities of ET scenarios and further approach the theoretical limits.

\begin{figure}[t]
\centering
\includegraphics[width = 0.9\linewidth]{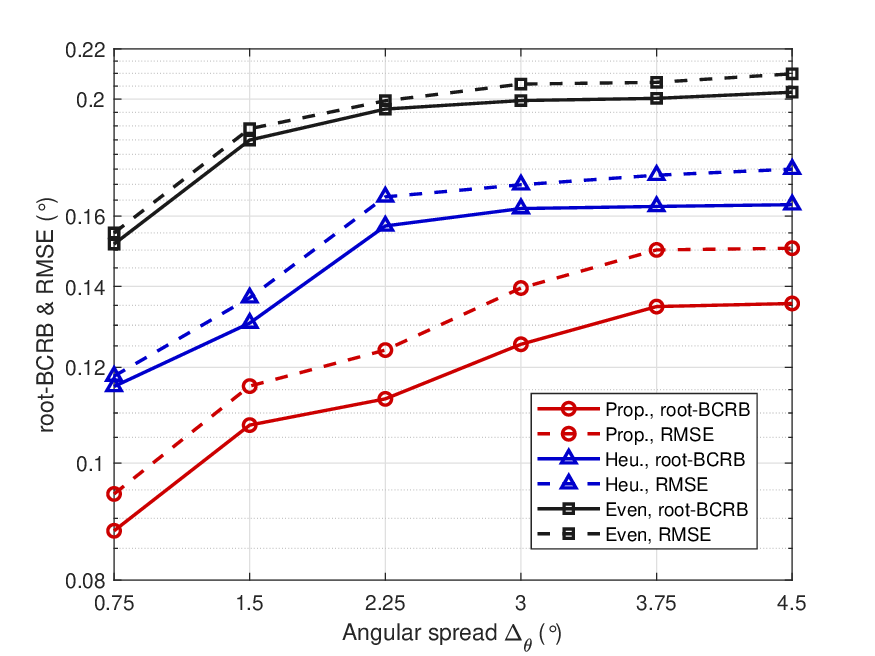}	
\caption{Root-BCRB and RMSE versus angular spread $\Delta_\theta$.}\label{fig:AS}\vspace{-0.3 cm}
\end{figure}

Finally, Fig.~\ref{fig:AS} illustrates the impact of angular spread on angle estimation performance. With the width of the angular bins fixed at $1.5^\circ$, an increase in angular spread results in more reflection points and additional RCS parameters, increasing estimation complexity and leading to higher estimation errors. Despite this, the proposed method consistently achieves lower estimation errors compared to other approaches, demonstrating its robustness and efficiency even as the angular spread increases. 

\vspace{-0.3 cm}
\section{Conclusion}
In this paper, we introduced a dynamic array partitioning strategy for monostatic ISAC systems to harness the available spatial degrees of freedom for enhanced sensing performance. We considered both multiple point-like and extended target scenarios, deriving the respective Bayesian CRB for each case assuming the availability of priors for the parameters of interest. Building on these derivations, we formulated a joint optimization problem that minimizes a weighted BCRB while simultaneously satisfying power budget, communication SINR, and partitioning requirements. To solve this problem, we proposed an alternating algorithm based on ADMM and SDR. A tailored MAP estimation approach was then developed for robust parameter recovery. Extensive simulation results demonstrated the pronounced advantages of dynamic partitioning over conventional fixed-partition arrays in terms of both theoretical BCRBs and empirical RMSE. Notably, strategies such as simply dividing the array into two halves or placing the receive antennas at the edges were shown to be inefficient, especially in complex sensing situations. 
\vspace{-0.3 cm}
\begin{appendices}
\section{Calculations for Gradient and Hessian Matrices}

In this appendix, we detail the derivations of the gradient and Hessian matrices used in Algorithm 2 for both the multiple point-like target and extended target scenarios. For the former, the gradient of $\mathcal{L}(\bm{\theta})$ with respect to $\theta_i$ is given by
\begin{align}
\label{eq:gradient}
\frac{\partial \mathcal{L}}{\partial \theta_i}  = [\bm{\Sigma}_\theta^{-1}]_{i,:} (\bm{\theta} - \bm{\mu}_\theta) - 2 \Re \big\{\mathbf{r}^H\widetilde{\mathbf{R}}_\text{n}^{-1}\dot{\mathbf{V}}_i\mathbf{D}^{-1}\mathbf{p}\big\},   
\end{align}
where $\mathbf{r} = \widetilde{\mathbf{y}}_\text{r}-\mathbf{VD}^{-1}\mathbf{p}$ and $\dot{\mathbf{V}}_i$ represents the partial derivation of $\mathbf{V}(\bm{\theta})$ with respect to $\theta_i$, given by $\dot{\mathbf{V}}_i = \frac{\partial\mathbf{V}}{\partial\theta_i} = \big[\mathbf{0},\ldots,\mathbf{0},\text{vec}\{\dot{\mathbf{H}}_i\mathbf{WS}\},\mathbf{0},\ldots,\mathbf{0}\big]$. 
The $(i,j)$-th element of the Hessian matrix is computed as 
\begin{equation}\begin{aligned}\label{eq:Hessian}
\frac{\partial^2 \mathcal{L}}{\partial\theta_i\partial\theta_j}& = [\bm{\Sigma}_\theta^{-1}]_{i,j} -2\Re \Big\{\frac{\partial\mathbf{r}^H}{\partial\theta_j}\widetilde{\mathbf{R}}_\text{n}^{-1}\dot{\mathbf{V}}_i\mathbf{D}^{-1}\mathbf{p}\Big\} \\
&\quad~  -2\Re \big\{\mathbf{r}^H\widetilde{\mathbf{R}}_\text{n}^{-1}\ddot{\mathbf{V}}_{i,j}\mathbf{D}^{-1}\mathbf{p}\big\}\\
&\quad~ +2\Re \Big\{\mathbf{r}^H\widetilde{\mathbf{R}}_\text{n}^{-1}\dot{\mathbf{V}}_{i}\mathbf{D}^{-1}\frac{\partial\mathbf{D}}{\partial\theta_j}\mathbf{D}^{-1}\mathbf{p}\Big\}\\
&\quad~ -2\Re \Big\{\mathbf{r}^H\widetilde{\mathbf{R}}_\text{n}^{-1}\dot{\mathbf{V}}_{i}\mathbf{D}^{-1}\frac{\partial\mathbf{p}}{\partial\theta_j}\Big\},
\end{aligned}\end{equation}
where we define the necessary intermediate derivatives 
\begin{subequations}\begin{align}
\frac{\partial\mathbf{r}}{\partial\theta_j} &= \mathbf{V}\mathbf{D}^{-1}\frac{\partial\mathbf{D}}{\partial\theta_j}\mathbf{D}^{-1}\mathbf{p}-\dot{\mathbf{V}}_j\mathbf{D}^{-1}\mathbf{p}  - \mathbf{V}\mathbf{D}^{-1}\frac{\partial\mathbf{p}}{\partial\theta_j},\\
\frac{\partial\mathbf{D}}{\partial\theta_j}&=\dot{\mathbf{V}}_j^H\widetilde{\mathbf{R}}_\text{n}^{-1} \mathbf{V} + \mathbf{V}^H\widetilde{\mathbf{R}}_\text{n}^{-1} \dot{\mathbf{V}}_j,\\
\frac{\partial\mathbf{p}}{\partial\theta_j}&= \dot{\mathbf{V}}_j^H\widetilde{\mathbf{R}}_\text{n}^{-1} \widetilde{\mathbf{y}}_\text{r},\\
\ddot{\mathbf{V}}_{i,i}&= \frac{\partial^2\mathbf{V}}{\partial^2\theta_i} = [\mathbf{0},\ldots,\mathbf{0},\text{vec}\{\ddot{\mathbf{H}}_i\mathbf{WS}\},\mathbf{0},\ldots,\mathbf{0}],\\
\ddot{\mathbf{V}}_{i,j}&= \mathbf{0},~\forall j\neq i.
\end{align}\end{subequations}
The second-order derivative $\ddot{\mathbf{H}}_i$ is computed based on the expressions for $\mathbf{H}_i$ and $\dot{\mathbf{H}}_i$ given in \eqref{eq:Ht Htd}. Specifically, 
\be
\ddot{\mathbf{H}}_i = (\mathbf{I}_N-\mathbf{A})\big[\ddot{\mathbf{h}}_i\mathbf{h}_i^T + 2\dot{\mathbf{h}}_i\dot{\mathbf{h}}_i^T +\mathbf{h}_i\ddot{\mathbf{h}}_i^T\big]\mathbf{A},
\ee
where $\ddot{\mathbf{h}}_i = \partial\dot{\mathbf{h}}_i/\partial\theta_i = \jmath\pi\sin\theta_i\mathbf{Q}\mathbf{h}_i-\pi^2\cos^2\theta_i\mathbf{Q}^2\mathbf{h}_i$.

For the ET scenario, the structure of the gradient and Hessian terms are identical to those in \eqref{eq:gradient} and \eqref{eq:Hessian}. The key distinction lies in how $\mathbf{V}$ depends on both the central angle $\theta_\text{c}$ and the angular spread $\Delta_\theta$. Consequently, the partial derivatives of $\mathbf{V}$ differ, leading to updated expressions for $\dot{\mathbf{V}}$ and $\ddot{\mathbf{V}}$. Once these derivatives are defined, they can be directly incorporated into the same gradient and Hessian expressions used for point-like targets, with only minimal changes to the associated computational steps. Thus, we only need to compute the following partial derivatives: 
\begin{subequations}\label{eq:Vij4et}\begin{align}
\dot{\mathbf{V}}_1 &\triangleq \frac{\partial\mathbf{V}}{\partial\theta_\text{c}} = \big[\ldots, \text{vec}\big\{\dot{\mathbf{H}}_i^{\theta_\text{c}}\mathbf{WS}\big\},\ldots\big],\\
\dot{\mathbf{V}}_2 &\triangleq \frac{\partial\mathbf{V}}{\partial\Delta_\theta} = \big[\ldots,\text{vec}\big\{\dot{\mathbf{H}}_i^{\Delta_\theta}\mathbf{WS}\big\},\ldots\big],\\
\ddot{\mathbf{V}}_{1,1} &\triangleq \frac{\partial^2\mathbf{V}}{\partial^2\theta_\text{c}} = \big[\ldots, \text{vec}\big\{\ddot{\mathbf{H}}_i^{\theta_\text{c}}\mathbf{WS}\big\},\ldots\big],\\
\ddot{\mathbf{V}}_{2,2} &\triangleq \frac{\partial^2\mathbf{V}}{\partial^2\Delta_\theta} = \big[\ldots,\text{vec}\big\{\ddot{\mathbf{H}}_i^{\Delta_\theta}\mathbf{WS}\big\},\ldots\big],\\
\ddot{\mathbf{V}}_{1,2} &\triangleq \frac{\partial^2\mathbf{V}}{\partial\theta_\text{c}\partial\Delta_\theta} = \big[\ldots,\text{vec}\big\{\ddot{\mathbf{H}}_i\mathbf{WS}\big\},\ldots\big],
\end{align}\end{subequations}
where we define
\begin{subequations}\begin{align}
\dot{\mathbf{H}}_i^{\theta_\text{c}} &\triangleq \frac{\partial\mathbf{H}_i}{\partial\theta_\text{c}}=-\jmath\pi\cos\theta_i(\mathbf{QH}_i+\mathbf{H}_i\mathbf{Q}),\\
\dot{\mathbf{H}}_i^{\Delta_\theta} &\triangleq \frac{\partial\mathbf{H}_i}{\partial\Delta_\theta}=-\jmath w_i\pi \cos\theta_i(\mathbf{QH}_i+\mathbf{H}_i\mathbf{Q}),\\
\ddot{\mathbf{H}}_i &\triangleq \frac{\partial^2\mathbf{H}_i}{\partial\theta_\text{c}\partial\Delta_\theta} = \jmath w_i\pi\sin\theta_i(\mathbf{QH}_i+\mathbf{H}_i\mathbf{Q})  \non\\
& \qquad\qquad\qquad  -\jmath\pi\cos\theta_i(\mathbf{Q}\dot{\mathbf{H}}_i^{\Delta_\theta}+\dot{\mathbf{H}}_i^{\Delta_\theta}\mathbf{Q}),\\
\ddot{\mathbf{H}}_i^{\theta_\text{c}} &\triangleq \frac{\partial^2\mathbf{H}_i}{\partial^2\theta_\text{c}} = \jmath\pi\sin\theta_i(\mathbf{QH}_i+\mathbf{H}_i\mathbf{Q}) \non\\
&\qquad\qquad\qquad -\jmath\pi\cos\theta_i(\mathbf{Q}\dot{\mathbf{H}}_i^{\theta_\text{c}}+\dot{\mathbf{H}}_i^{\theta_\text{c}}\mathbf{Q}),\\
\ddot{\mathbf{H}}_i^{\Delta_\theta} &\triangleq \frac{\partial^2\mathbf{H}_i}{\partial^2\Delta_\theta} = \jmath w_i^2\pi\sin\theta_i(\mathbf{QH}_i+\mathbf{H}_i\mathbf{Q}) \non\\
&\qquad\qquad\qquad -\jmath w_i\pi\cos\theta_i(\mathbf{Q}\dot{\mathbf{H}}_i^{\Delta_\theta}+\dot{\mathbf{H}}_i^{\Delta_\theta}\mathbf{Q}).
\end{align}\end{subequations}
Substituting the definitions of $\dot{\mathbf{V}}_i$ and $\mathbf{V}_{i,j}$ from \eqref{eq:Vij4et} into \eqref{eq:gradient} and \eqref{eq:Hessian}, the corresponding gradient and Hessian matrices for the ET estimation problem can be readily obtained. 

\end{appendices}

\end{document}